\documentclass[%
 aip,
 amsmath,amssymb,
 reprint,%
 citeautoscript,
]{revtex4-1}

\usepackage{graphicx}
\usepackage{bm}
\usepackage[utf8]{inputenc}
\usepackage[T1]{fontenc}
\usepackage{mathptmx}
\usepackage{siunitx}
\usepackage{xcolor}
\usepackage{mathtools} 
\newlength{\CapLen}
\AtBeginDocument{\settoheight{\CapLen}{K}} 
\newcommand{\Kappa}{\resizebox{!}{\CapLen}{$\kappa$}}

\newcommand\footnoteref[1]{\protected@xdef\@thefnmark{\ref{#1}}\@footnotemark}
\usepackage{longtable}
\usepackage{tablefootnote}

\begin{document}
\title{Improved accuracy and transferability of molecular-orbital-based machine learning: Organics, transition-metal complexes, non-covalent interactions, and transition states} 
\author{Tamara Husch}
\author{Jiace Sun}
\author{Lixue Cheng}
\author{Sebastian J. R. Lee}
\author{Thomas F. Miller III}
\email{tfm@caltech.edu} 
\affiliation{%
Division of Chemistry and Chemical Engineering, California Institute of Technology, Pasadena, CA 91125, USA
}%

\date{\today}

\begin{abstract}

Molecular-orbital-based machine learning (MOB-ML) provides a general framework for the prediction of accurate correlation energies at the cost of obtaining molecular orbitals.
The application of Nesbet's theorem makes it possible to recast a typical extrapolation task, training on correlation energies for small molecules and predicting correlation energies for large molecules, into an interpolation task based on the properties of orbital pairs. 
We demonstrate the importance of preserving physical constraints, including invariance conditions and size consistency, when generating the input for the machine learning model. 
Numerical improvements are demonstrated for different data sets covering total and relative energies for thermally accessible organic and transition-metal containing molecules, non-covalent interactions, and transition-state energies.
MOB-ML requires training data from only 1\% of the QM7b-T data set (i.e., only 70 organic molecules with seven and fewer heavy atoms) to predict the total energy of the remaining 99\% of this data set with sub-kcal/mol accuracy. 
This MOB-ML model is significantly more accurate than other methods when transferred to a data set comprised of thirteen heavy atom molecules, exhibiting no loss of accuracy on a size intensive (i.e., per-electron) basis.
It is shown that MOB-ML also works well for extrapolating to transition-state structures, predicting the barrier region for malonaldehyde intramolecular proton-transfer to within 0.35~kcal/mol when only trained on reactant/product-like structures. 
Finally, the use of the Gaussian process variance enables an active learning strategy for extending MOB-ML model to new regions of chemical space with minimal effort.
We demonstrate this active learning strategy by extending a QM7b-T model to describe non-covalent interactions in the protein backbone-backbone interaction data set to an accuracy of 0.28~kcal/mol.
\end{abstract}

\maketitle

\section{Introduction}

The calculation of accurate potential energies of molecules and materials at affordable cost is at the heart of computational chemistry.
While state-of-the-art \textit{ab initio} electronic structure theories can yield highly accurate results, they are computationally too expensive for routine applications.
Density functional theory (DFT) is computationally cheaper and has thus enjoyed widespread applicability.
However, DFT is hindered by a lack of systematic improvability and from an uncertain quality for many applications. 

In recent years, a variety of machine learning approaches has emerged which promise to mitigate the cost of highly accurate electronic structure methods while preserving accuracy. \cite{bartok_gaussian_2010, rupp_fast_2012,hansen_assessment_2013,ramakrishnan_big_2015, behler_perspective_2016, brockherde_bypassing_2017, schutt_schnet_2017, schutt_quantum-chemical_2017,  smith_ani-1_2017, mcgibbon_improving_2017, collins_constant_2018, fujikake_gaussian_2018, lubbers_hierarchical_2018,  nguyen_comparison_2018, welborn_transferability_2018, wu_moleculenet_2018, yao_tensormol-01_2018, cheng_universal_2019, cheng_regression_2019, christensen_operators_2019, grisafi_transferable_2019, smith_approaching_2019, unke_physnet_2019,
fabrizio_machine_2020, chen_ground_2020, christensen_fchl_2020, dick_machine_2020, liu_transferable_2020,  qiao_orbnet_2020,manzhos_machine_2020}
While these machine learning methods share similar goals, they differ in the representation of the molecules and in the machine learning methodology itself. 
Here, we will focus on the molecular-orbital-based machine learning (MOB-ML) approach. \cite{welborn_transferability_2018, cheng_universal_2019, cheng_regression_2019}
The defining feature of MOB-ML is its framing of learning highly accurate correlation energies as learning a sum of orbital pair correlation energies. These orbital pair correlation energies can be individually  regressed with respect to a feature vector representing the interaction of the molecular orbital pairs.
Without approximation, it can be shown that such pair correlation energies add up to the correct total correlation energy for single-reference wave function methods. 
Phrasing the learning problem in this manner has the  advantage that a given pair correlation energy, and, hence, a given feature vector, is independent of molecular size (after a certain size threshold has been reached) because of the inherent spatial locality of dynamic electron correlation.
Consequently, operating in such an orbital pair interaction framework converts the general extrapolation task of training on small molecules and predicting on large molecule into an interpolation task of training on orbital pairs in a small molecule and predicting on the same orbital pairs in a large molecule.

In this work, we address challenges introduced by operating in a vectorized molecular orbital pair interaction framework (Section~\ref{sec:theory}). We  show how changes to the feature design affect the performance and transferability of MOB-ML models within the same molecular family (Section~\ref{subsec:sec1}) and across molecular families (Sections~\ref{subsec:sec2}-\ref{subsec:sec3}). We probe these effects on relative- and total-energy predictions for organic and transition-metal containing molecules, and we investigate the applicability of MOB-ML to transition-state structures and non-covalent interactions.

\section{Theory}
\label{sec:theory}

MOB-ML predicts correlation energies based on information from the molecular orbitals. \cite{welborn_transferability_2018, cheng_universal_2019, cheng_regression_2019}
The correlation energy $E^\text{corr}$ in the current study is defined as the difference between the true total electronic energy and the Hartree--Fock (HF) energy for a given basis set. Without approximation, the correlation energy is expressed as a sum over correlation energy contributions from pairs of occupied orbitals $i$ and $j$, \cite{nesbet_brueckners_1958}
\begin{equation}
     E^\text{corr} = \sum_{ij} \epsilon_{ij}.
\end{equation}
Electronic structure theories offer different ways of approximating these pair correlation energies.
For example, the second-order M{\o}ller-Plesset perturbation theory  (MP2) correlation energy is \cite{moller_note_1934} 
\begin{equation}
     \epsilon_{ij}^\text{MP2} = \sum_{ab} \frac{\left<ia||jb\right>^2}{F_{aa}+F_{bb} - F_{ii} - F_{jj}}, 
\end{equation}
where $a,b$ denote virtual orbitals, $F$ the Fock matrix in the molecular orbital basis, and $\left<ia||jb\right>$ the anti-symmetrized exchange integral.
We denote a general repulsion integral over the spatial coordinates $\mathbf{x}_1, \mathbf{x}_2$ of molecular orbitals $p,q,m,n$ following the chemist's notation as 
\begin{equation}
\begin{split}
[\Kappa^{pq}]_{mn} &= \left< pq | mn \right> \\&= \int d \mathbf{x}_1 d \mathbf{x}_2 p(\mathbf{x}_1)^* q(\mathbf{x}_1) \frac{1}{|\mathbf{x}_1 - \mathbf{x}_2|} m(\mathbf{x}_2)^* n(\mathbf{x}_2).
\end{split}
\end{equation}
The evaluation of correlation energies with post-HF methods like MP2 or coupled-cluster theory (including CCSD(T)) involves  computations that  exceed the cost of HF theory by orders of magnitude.
By contrast, MOB-ML predicts the correlation energy at negligible cost by machine-learning the map  
\begin{equation}
    \epsilon_{ij} \approx \epsilon^\text{ML}(\mathbf{f}_{ij}),
\end{equation}
where $\mathbf{f}_{ij}$ denotes the feature vector into which information on the molecular orbitals is compiled. 

Following our previous work,\cite{cheng_universal_2019} we define a canonical order of the orbitals $i$ and $j$ by rotating them into gerade and ungerade combinations (see Eq.~(7) in Ref.~\onlinecite{cheng_universal_2019}), creating the rotated orbitals $\tilde{i}$ and $\tilde{j}$. 
The feature vector $\mathbf{f}_{ij}$ assembles information on the molecular orbital interactions: (i)  Orbital energies of the valence-occupied and valence-virtual orbitals $F_{pp}$, (ii) 
mean-field interaction energy of valence-occupied and valence-occupied orbitals and of valence-virtual and valence-virtual orbitals $F_{pq}$, 
(iii) Coulomb interaction of valence-occupied and valence-occupied orbitals, of valence-occupied and valence-virtual orbitals, and valence-virtual and valence-virtual orbitals $[\Kappa^{pp}]_{qq}$, and
(iv) exchange interaction of valence-occupied and valence-occupied orbitals, of valence-occupied and valence-virtual orbitals, and valence-virtual and valence-virtual orbitals $[\Kappa^{pq}]_{pq}$.
We note that all of these pieces of information enter either the MP2 or the MP3 correlation energy expressions, which helps to motivate their value within our machine learning framework.
We remove  repetitive information from the feature vector and  separate the learning problem into the cases where (i) $i\ne j$ where we employ the feature vector as defined in Eq.~(\ref{eq:off-diag_f}) and (ii) $i=j$ where we employ the feature vector as defined in Eq.~(\ref{eq:diag_f}),
\onecolumngrid
\begin{equation}
\label{eq:off-diag_f}
\begin{split}
    \mathbf{f}_{ij} =& \{ \{F_{\tilde{i}\tilde{i}}, F_{\tilde{i}\tilde{j}}, F_{\tilde{j}\tilde{j}}\}, \{F_{\tilde{i}k}\}, \{F_{\tilde{j}k}\}, \{F_{ab}\}, \\
    &   \{[\Kappa^{\tilde{i}\tilde{i}}]_{\tilde{i}\tilde{i}}, [\Kappa^{\tilde{i}\tilde{i}}]_{\tilde{j}\tilde{j}}, [\Kappa^{\tilde{j}\tilde{j}}]_{\tilde{j}\tilde{j}}\}, \{[\Kappa^{\tilde{i}\tilde{i}}]_{kk}\}, \{[\Kappa^{\tilde{j}\tilde{j}}]_{kk}\}, \{[\Kappa^{\tilde{i}\tilde{i}}]_{aa}\}, \{[\Kappa^{\tilde{j}\tilde{j}}]_{aa}\}, \{[\Kappa^{aa}]_{bb}\} , \\
    &  \{[\Kappa^{\tilde{i}\tilde{j}}]_{\tilde{i}\tilde{j}}\}, \{[\Kappa^{\tilde{i}k}]_{\tilde{i}k}\}, \{[\Kappa^{\tilde{j}k}]_{\tilde{j}k}\}, \{[\Kappa^{\tilde{i}a}]_{\tilde{i}a}\}, \{[\Kappa^{\tilde{j}a}]_{\tilde{j}a}\}, \{[\Kappa^{ab}]_{ab}\} \}, \\
\end{split}
\end{equation}{}
\begin{equation}
\label{eq:diag_f}
\begin{split}
    \mathbf{f}_{i} =& \{ F_{ii}, \{F_{ik}\}, \{F_{ab}\},  [\Kappa^{ii}]_{ii}, \{[\Kappa^{ii}]_{kk}\},  \{[\Kappa^{ii}]_{aa}\}, \{[\Kappa^{aa}]_{bb}\},  \{[\Kappa^{ik}]_{ik}\}, \{[\Kappa^{ia}]_{ia}\}, \{[\Kappa^{ab}]_{ab}\} \}. \\
\end{split}
\end{equation}{}
\twocolumngrid
\noindent
Here, the index $k$ denotes an occupied orbital other than $i$ and $j$.  
For blocks in the feature vector that include more than one element, we specify a canonical order of the feature vector elements. 
In our previous work,\cite{cheng_universal_2019} this order was given by the sum of the Euclidean distances between the centroids of orbital $\tilde{i}$ and $p$ and between the centroids of orbital $\tilde{j}$ and $p$.
In the current work, we introduce a different strategy to sort the feature vector elements (Section~\ref{sec:sorting}), we modify the protocol with which we obtain the feature vector elements associated with $\tilde{i}, \tilde{j}$ (Section~\ref{sec:invariance}), and we revise our feature vector elements to ensure size consistency (Section~\ref{sec:sizeconsistency}).
We provide a conceptual description of the changes to the feature set below and we give the full definition of the feature vector elements and the criteria according to which the feature elements are ordered in Tables~S3--S6 in the Supporting information.

\subsection{Defining importance of feature vector elements}
\label{sec:sorting}
Careful ordering of the elements of the feature vector blocks in necessary in the current work  because Gaussian process regression (GPR) is  sensitive to permutation of the feature vector elements. 
Furthermore, the application of a Gaussian process requires that the feature vectors be of fixed length. \cite{rasmussen_gaussian_2006} 

Given the near-sighted nature of dynamical electron correlation, it is expected that only a limited number of orbital-pair interactions are important to predict the pair correlation energy with MOB-ML.  
To construct the fixed-length feature vector, a cutoff criterion must be introduced.\cite{welborn_transferability_2018} 
For some feature vector elements, a robust definition of importance is straight-forward.
The spatial distance between the orbital centroids $i$ and $a$ is, for example, a reliable proxy for the importance of the feature vector elements $\{[\Kappa^{ii}]_{aa}\}$ of the feature vector $ \mathbf{f}_{i}$.
However, the definition of importance is less straightforward for feature vector elements that involve more than two indices. 
The most prominent example is the $\{[\Kappa^{ab}]_{ab}\}$ feature vector block of $\mathbf{f}_{ij}$, which contains the exchange integrals between the valence-virtual orbitals $a$ and $b$ and which should be sorted with respect to the importance of these integrals for the prediction of the pair correlation energy $\epsilon_{ij}$. 
It is non-trivial to define a spatial metric which defines the importance of the feature vector elements $\{[\Kappa^{ab}]_{ab}\}$ to predict the pair correlation energy $\epsilon_{ij}$; instead, we employ the the MP3 approximation for the pair correlation energy, 
\begin{equation}
\label{eq:mp3}
\begin{split} 
    \epsilon_{ij}^\text{MP3} &= \frac{1}{8} \sum_{abcd} 
    \left(t_{ij}^{ab}\right)^* \left<ab||cd\right> t_{ij}^{cd} +
    \frac{1}{8} \sum_{klab} 
     \left(t_{ij}^{ab}\right)^* \left<kl||ij\right> t_{kl}^{ab} 
    \\&-
    \sum_{kabc} 
     \left(t_{ij}^{ab}\right)^* \left<kb||ic\right> t_{kj}^{ac}, \\
\end{split}
\end{equation}
where $t_{ij}^{ab}$ denotes the T-amplitude. 
Although we operate in a local molecular orbital basis,  the canonical formulae are used to define the importance criterion; if we consider orbital localization as a perturbation (as in Kapuy--M{\o}ller--Plesset theory \cite{kapuy_application_1983}), the canonical expression is the leading order term. 
The term we seek to attach an importance to, $\{[\Kappa^{ab}]_{ab}\}$, appears in the first term of  Eq.~(\ref{eq:mp3}) and all integrals necessary to compute this term are readily available as (a combination of) other feature elements, i.e., we do not incur any additional significant computational cost to obtain the importance of the feature vector elements. 

The way in which we determine the importance of the $\{[\Kappa^{ab}]_{ab}\}$ elements here is an example of a more general strategy that we employ, in which the importance is assigned according to the lowest-order perturbation theory in which the features first appear in. 
Similar considerations have to be made for each feature vector block, all of which are specified in detail in Tables~S3 and S4 in the Supporting Information. 

\subsection{Orbital-index permutation invariance}
\label{sec:invariance}

The Fock, Coulomb, and exchange matrix elements that comprise MOB features are naturally invariant to rotation and translation of the molecule.
However, some care is needed to ensure that these invariances are not lost in the construction of symmetrized MOB features. 
In particular, rotating the valence-occupied orbitals into gerade and  ungerade combinations leads to an orbital-index permutation variance for energetically degenerate orbitals $i,j$ because the sign of the feature vector elements $M_{\tilde{j}p}$,
\begin{equation}
 M_{\tilde{j}p} = \frac{1}{\sqrt{2}} M_{ip} - \frac{1}{\sqrt{2}} M_{jp},
\end{equation}
depends on the arbitrary assignment of the indices $i$ and $j$. 
To rectify this issue, we include the absolute value of the generic feature vector element $M$ in the feature vector instead of the signed value,
\begin{equation}
 M_{\tilde{j}p} = \left| \frac{1}{\sqrt{2}} M_{ip} - \frac{1}{\sqrt{2}} M_{jp} \right|,
\end{equation}
where $M_{\tilde{j}p}$ may be $F_{\tilde{j}p}$, $[\Kappa^{\tilde{j}\tilde{j}}]_{pp}$, or $[\Kappa^{\tilde{j}p}]_{\tilde{j}p}$.
The corresponding equation, 
\begin{equation}
 M_{\tilde{j}p} = \frac{1}{\sqrt{2}} M_{ip} + \frac{1}{\sqrt{2}} M_{jp},
\end{equation}
is already orbital-index permutation invariant because we chose $M_{pq}$ ($p\ne q$) to be positive. \cite{cheng_universal_2019} 

\subsection{Size consistency}
\label{sec:sizeconsistency}

Size consistency is the formal property by which the energy of two isolated molecules equals the sum of their dimer upon infinite separation.\cite{bartlett_many-body_1981, SzaboSizeConsistent}
In the context of MOB-ML, satisfaction of this property requires that the contributions from the diagonal feature vectors are not affected by distant, non-interacting molecules and that 
\begin{equation}
    \epsilon^{ML}(\mathbf{f}_{ij})=0 \text{ for } r_{ij} = \infty
\end{equation}
for contributions from the off-diagonal feature vectors.
To ensure that MOB-ML exhibits size-consistency without the need for explicit training on the dimeric species, the following modifications to the feature vectors are made. 

\paragraph{Diagonal feature vector.}
The feature vector as defined in Eq.~(\ref{eq:diag_f}) contains three blocks whose elements are independent of orbital $i$, $\{F_{ab}\}$, $\{[\Kappa^{aa}]_{bb}\}$, and $\{[\Kappa^{ab}]_{ab}\}$. The magnitude of these feature vector elements does not decay with an increasing distance between orbital $i$ localized on molecule $I$ and an orbital (for example, $a$) localized on molecule $J$.
To address this issue, we multiply these feature vector elements by their estimated importance (see Section~\ref{sec:sorting}) so that they decay smoothly to zero.
The other feature vector elements decay to zero when the involved orbitals are non-interacting albeit at different rates; we take the cube of feature vector elements of the type $\{[\Kappa^{pp}]_{qq}\}$ to achieve a similar decay rate for all feature vector elements in the short- to medium-range which facilitates machine learning.

\paragraph{Off-diagonal feature vector.}
We modify the off-diagonal feature vector such that $\mathbf{f}_{ij}=\mathbf{0} \text{ for }r_{ij} = \infty$ by first applying the newly introduced  changes  for $\mathbf{f}_{i}$ also for $\mathbf{f}_{ij}$. 
Further action is needed for the off-diagonal case because many feature vector elements do not decay to zero when the distance between $i$ and $j$ is large due to rotation of the orbitals into a gerade and an ungerade combination, e.g., $F_{\tilde{i}k}=  \left| \frac{1}{\sqrt{2}} F_{ik} + \frac{1}{\sqrt{2}} F_{jk} \right| = \left| \frac{1}{\sqrt{2}} F_{ik} \right| \text{for } r_{ij}=\infty, r_{jk}=\infty$.
As a remedy, we apply a damping function of the form $ \frac{1}{1+\frac{1}{6}(r_{ij}/r_0)^6}$ to each feature vector element.
The form of this damping function is inspired by the semi-classical limit of the MP2 expression as it is also used for semi-classical dispersion corrections. \cite{grimme_dispersion_2016} 
The damping radius, $r_0$, needs to be sufficiently large as to not interfere with machine learning at small $r_{ij}$. 
If a damping radius close to zero would be chosen, all off-diagonal feature vectors would be zero which nullifies the information content; however,  the damping radius $r_0$ also should not be too large as size-consistency has to be fully learned until the off-diagonal feature vector is fully damped to zero. Therefore, we employ a damping radius in the intermediate-distance regime and we empirically found $r_0 = 5.0$~Bohr to work well.

Lastly, we enforce that $\epsilon^{ML}(\mathbf{0})=0$.
The MOB features are engineered to respect this limit and would, for example, in a linear regression with a zero intercept trivially predict a zero-valued pair correlation energy without any additional training.
However, the Gaussian process regression we apply in this work does not trivially yield a zero-valued pair correlation energy for a zero-valued feature vector.
In the case that a training set does not include examples of zero-valued feature vectors, we need to include zero-valued feature vectors and zero-valued pair correlation energies in training to ensure that $\epsilon^{ML}(\mathbf{0})=0$. 
For no model trained in the current study were more than 5\% zero-valued feature vectors included.

The resulting MOB-ML model leads to size consistent energy predictions to the degree to which the underlying MO generation is. 
It is not required that the dimer is explicitly part of training the MOB-ML model to obtain this result. 
The detailed definition of each feature vector block is summarized in Tables~S5 and S6. We apply the feature set defined in Tables~S5 and S6 consistently in this work.

\section{Computational details}

We present results for five different data sets: 
(i) a series of alkane molecules, 
(ii) the potential energy surface of the malonaldehyde molecule, 
(iii) a thermalized version of the QM7b and the GDB13 data set (i.e., QM7b-T and GDB13-T), \cite{cheng_thermalized_2019}
(iv) a set of backbone-backbone interactions (BBI) \cite{burns_biofragment_2017}, and
(v) a thermalized version of a subset of mononuclear, octahedral transition metal complexes put forward by Kulik and co-workers \cite{nandy_strategies_2018}.
We refer to the Supporting Information Section II for a description how the structures were obtained or generated. 
All generated structures are available in Ref.~\onlinecite{caltech_data}.

The features for all structures were generated with the \textsc{entos qcore} \cite{manby_entos_2019} package.
The feature generation is based on a HF calculation applying a cc-pVTZ\cite{dunning_gaussian_1989} basis for the elements H, C, N, O, S, and Cl. We apply a def2-TZVP basis set \cite{weigend_balanced_2005} for all transition metals.  
The HF calculations were accelerated with density fitting for which we applied the corresponding cc-pVTZ-JKFIT\cite{weigend_fully_2002} and def2-TZVP-JKFIT \cite{weigend_accurate_2006} density fitting bases.
Subsequently, we localized the valence-occupied and the valence-virtual molecular orbitals with the Boys--Foster localization scheme \cite{boys_construction_1960,foster_canonical_1960} or with the intrinsic bond orbital (IBO) localization scheme \cite{knizia_intrinsic_2013}.
We implemented a scheme to localize the valence-virtual orbitals with respect to the Boys--Foster function (for details on this implementation, see Section II in the Supplementary Information).
We applied the Boys--Foster localization scheme for the data sets (i), (iii), (iv), and (v) for valence-occupied and valence-virtual molecular orbitals.
IBO localization for valence-occupied and valence-virtual molecular orbitals led to better results for data set (ii).

The resulting orbitals are imported into the Molpro 2018.0 \cite{werner_molpro_2018,werner_molpro_2020} package via the \texttt{matrop} functionality to generate the non-canonical MP2 \cite{werner_fast_2003} or CCSD(T) \cite{scuseria_comparison_1990, hampel_local_1996, schutz_local_2000} pair correlation energies  with the same orbitals we applied for the feature generation. 
These calculations are accelerated with the resolution of the identity approximation. 
The frozen-core approximation is invoked for all correlated calculations. 

We follow the machine learning protocol outlined in   previous work \cite{cheng_universal_2019} to train the MOB-ML models. 
In a first step, we perform MOB feature selection by evaluating the mean decrease of accuracy in a random forest regression in the \textsc{Scikit-learn} v0.22.0 package \cite{pedregosa_scikit-learn_2011}.
We then regress the diagonal and off-diagonal pair correlation energies separately with respect to the selected features in the \textsc{GPy} 1.9.6 software package. \cite{gpy_gpy_2012} 
We employ the Mat\'ern 5/2 kernel with white noise regularization\cite{rasmussen_gaussian_2006}.
We minimize the negative log marginal likelihood objective with respect to the kernel hyperparameters with a scaled conjugate gradient scheme for 100 steps and then apply the BFGS algorithm until full convergence. 
As indicted in the results, both random-sampling and active-learning strategies were employed for the selection of molecules in the training data sets. 
In the active-learning strategy, we use a previously trained MOB-ML model to evaluate the Gaussian process variance for each molecule, 
and then include the points with the highest variance in the training data set, as outlined in Ref.~\onlinecite{shapeev_active_2020}.
To estimate the Gaussian process variance for each molecule, it was assumed the variances per molecular orbital pair are mutually independent.

\section{Results}

\subsection{Transferability within a molecular family}
\label{subsec:sec1}

\begin{figure}[htbp]
\includegraphics[width=\columnwidth]{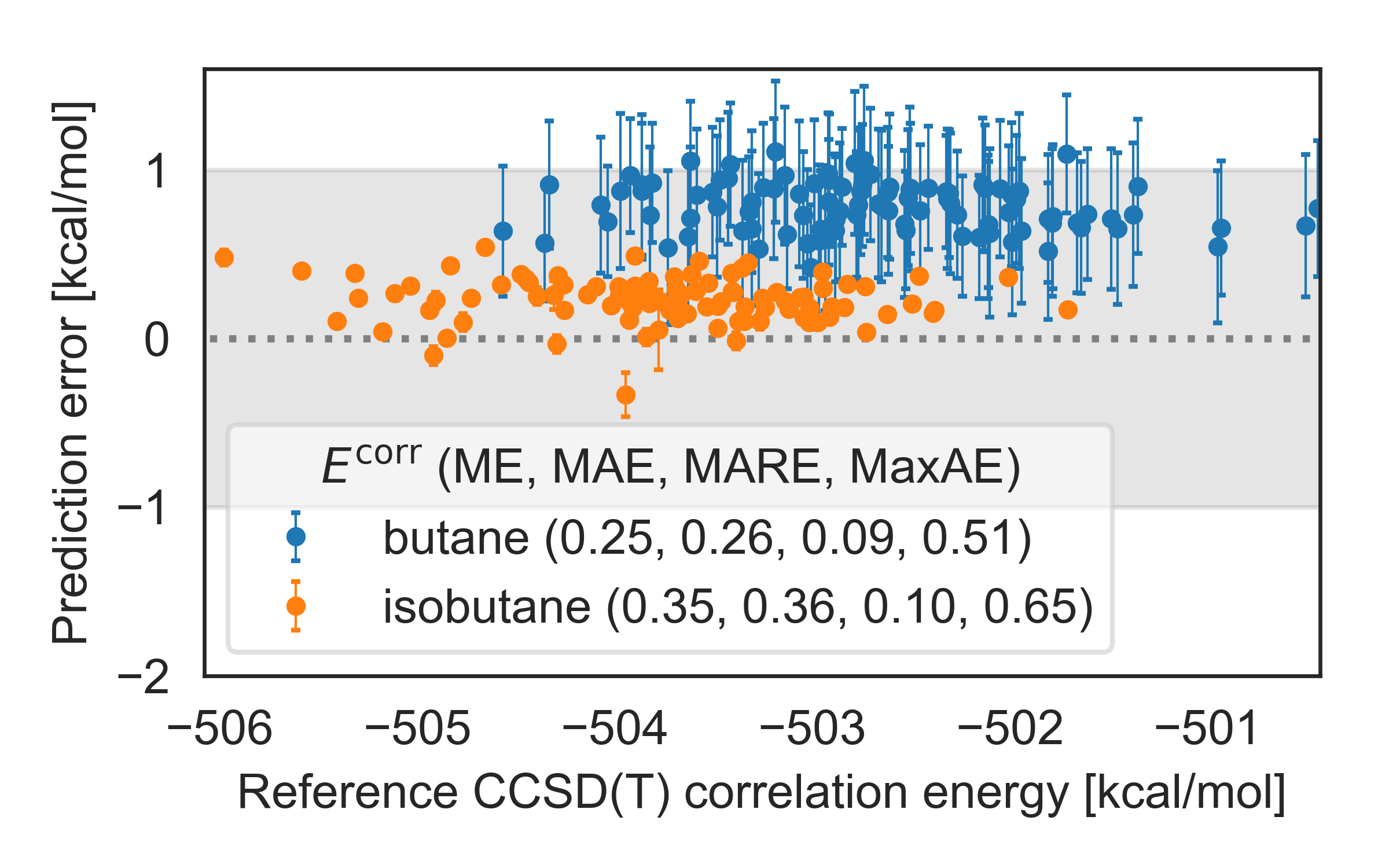}
\caption{
Errors in the predicted correlation energies with respect to the CCSD(T) reference values for butane and isobutane. The bar attached to each prediction error indicates the associated Gaussian process variance. The MOB-ML model used for these predictions was trained on 50 ethane and 20 propane molecules. The gray shaded area corresponds to the region where the error is smaller than chemical accuracy (1~kcal/mol). 
}
\label{figure:figure1}
\end{figure}

We first examine the effect of the feature vector generation strategy on the transferability of MOB-ML models within a molecular family.
To this end, we revisit our alkane data set \cite{cheng_universal_2019} which contains 1000 ethane and 1000 propane geometries as well as 100 butane and 100 isobutane geometries.
We perform the transferability test outlined in Ref.~\onlinecite{cheng_universal_2019}, i.e., training a MOB-ML model on correlation energies for 50 randomly chosen ethane geometries and 20 randomly chosen propane geometries to predict the correlation energies for the 100 butane and 100 isobutane geometries (see Figure~\ref{figure:figure1}).
This transferability test was repeated with 10000 different training data sets (each consisting of data for 50 ethane molecules and 20 propane molecules) to assess the training set dependence of the MOB-ML models.
As suggested in Ref.~\onlinecite{chen_ground_2020}, we consider various performance metrics to assess the prediction accuracy of the MOB-ML models: (i) the mean error (ME, Eq.~(S3)), (ii) the mean absolute error (MAE, Eq.~(S4)), (iii) the maximum absolute error (MaxAE, Eq.~(S5)), and (iv) the mean absolute relative error (MARE, Eq~(S6)) which applies a global shift setting the mean error to zero. 
We report the minimum, peak, and maximum encountered MAREs in Table~\ref{tab:table1} alongside literature values obtained in our previous work \cite{cheng_universal_2019}, by Dick \textit{et al.}, \cite{dick_machine_2020} and by Chen \textit{et al}. \cite{chen_ground_2020} The MEs, MAEs, and MaxAEs are reported in Figure~S1.
\begin{center}
\begin{table}[htbp]
    \centering
    \begin{tabular}{p{1.6cm}p{2.15cm}p{0.6cm}p{0.6cm}p{0.6cm}p{0.6cm}p{0.6cm}p{0.6cm}}
         \hline \hline 
         Method & Feature set & \multicolumn{6}{c}{MARE} \\
                &             & \multicolumn{3}{c}{Butane} & \multicolumn{3}{c}{Isobutane}\\
                & & min & peak & max & min & peak & max  \\
         \hline 
         NeuralXC\cite{dick_machine_2020} & --- & & 0.15 & & & 0.14 & \\
         DeepHF\cite{chen_ground_2020} & --- & 0.06& 0.11 & 0.43 & 0.07 & 0.13 & 0.53 \\
         \hline 
         MOB-ML & Ref.~\onlinecite{cheng_universal_2019} & & 0.20 & & & 0.21 & \\
         & \textbf{this work } & \textbf{0.06} & \textbf{0.11} & \textbf{0.19 } & \textbf{0.06} & \textbf{0.10} & \textbf{0.19}\\
         \hline \hline 
    \end{tabular}
    \caption{Comparison of the minimum, peak, and maximum  mean absolute error after global shift (MARE) in~kcal/mol for the prediction of CCSD(T) correlation energies for butane and isobutane obtained with different methods.}
    \label{tab:table1}
\end{table}
\end{center}
In general, MOB-ML as well as NeuralXC\cite{dick_machine_2020} and DeepHF\cite{chen_ground_2020} produce MAREs well below chemical accuracy for correlation energies of butane and isobutane when trained on correlation energies of ethane and propane.
Updating the feature vector generation strategy for MOB-ML results in the best peak MAREs for butane as well as for isobutane which are 0.11~kcal/mol and 0.10~kcal/mol, respectively.
As in our previous work, \cite{cheng_universal_2019} we note that the total correlation energy predictions may be shifted  with respect to the reference data so that the MEs for MOB-ML range from $-0.92$ to 2.70~kcal/mol for butane and from $-0.18$ to 1.02~kcal/mol for isobutane (see also Figure~S1). 
This shift is strongly training-set dependent, which was also observed for results obtained with DeepHF \cite{chen_ground_2020}. 

The results highlight that this is an extrapolative transferability test. 
A considerable advantage of applying GPR in practice is that each prediction is accompanied by a Gaussian process variance which, in this case, indicates that we are in an extrapolative regime (see Figure~\ref{figure:figure1}). 
Extrapolations might be associated with quality degradation which we see, most prominently, for the mean error in butane.
By contrast, other machine learning approaches like neural networks are less clear in terms of whether the predictions are in an interpolative or extrapolative regime.\cite{hirschfeld_uncertainty_2020} 
By including the butane molecule with the largest variance in the training set (which then consists of 50 ethane, 20 propane, and 1 butane geometries) we  reduce the ME from 0.78 to 0.25, MAE from 0.78 to 0.26, MaxAE from 1.11 to 0.51, and the MARE from 0.11 to 0.09~kcal/mol for butane (see Figure~S2). 
These results directly illustrate that MOB-ML can be systematically improved by including training data that is more similar to the test data; the improved confidence of the prediction is then also directly reflected in the associated Gaussian process variances. 

\begin{center}
\begin{figure}[htbp]
\includegraphics[width=\textwidth]{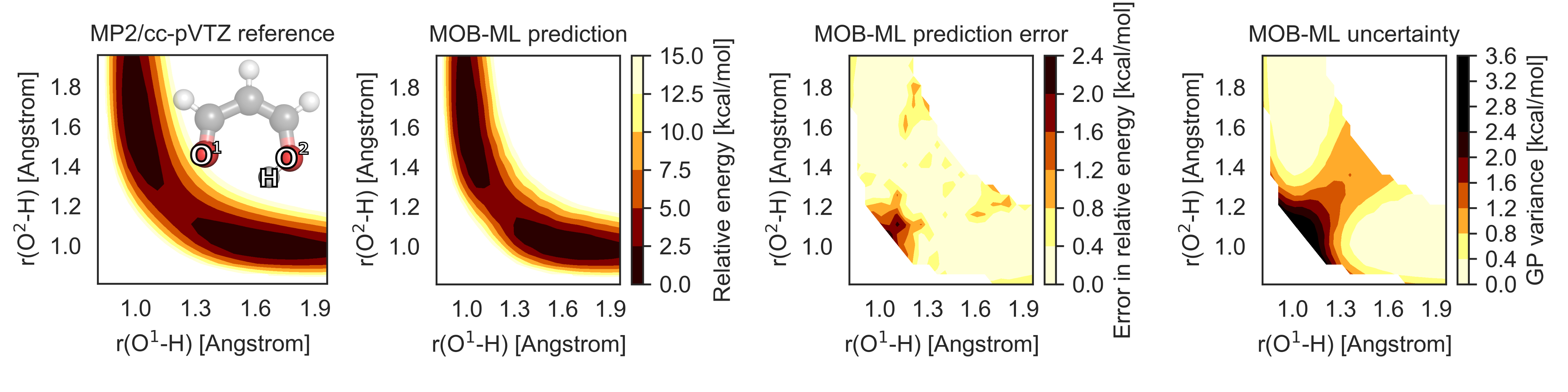}
\caption{
Relative energies obtained with MP2/cc-pVTZ (left panel), relative energies predicted with MOB-ML (middle left panel), the difference between the MOB-ML prediction and the reference data (middle right panel), and the Gaussian process variance (right panel) for the proton transfer in malonaldehyde as a function of the distance of the proton from the two oxygen atoms. 
}
\label{figure:figure2}
\end{figure}
\end{center}

As a second example, we examine the transferability of a MOB-ML model trained within a basin of a potential energy surface to the transition-state region of the same potential energy surface.
We chose malonaldehyde for this case study as it has also been explored in previous machine learning studies \cite{brockherde_bypassing_2017}.
We train a MOB-ML model on 50 thermalized malonaldehyde structures which all have the property that $d$(O$^1$--H) + $d$(O$^2$--H) > 0.4~\AA\ (where $d$ denotes the distance between the two nuclei) which ensures that we are sampling from the basins.
We then apply this trained model to predict the correlation energies for a potential energy surface mapping out the hydrogen transfer between the two oxygen atoms (see Figure~\ref{figure:figure2}). 
MOB-ML produces an accurate potential energy surface for the hydrogen transfer in malonaldehyde only from information on the basins (compare left and middle left panel of Figure~\ref{figure:figure2}). 
The highest encountered errors on the minimum potential energy path are smaller than 1.0~kcal/mol.
Unsurprisingly, the predicted minimum energy structure ($d$(O$^1$--H) = 1.00~\AA, $d$(O$^2$--H) = 1.63~\AA) is very similar to the reference minimum energy structure ($d$(O$^1$--H) = 1.00~\AA, $d$(O$^2$--H) = 1.64~\AA). 
Strikingly, the predicted energy of 2.65~kcal/mol at the saddle point at $d$(O$^1$--H) = $d$(O$^2$--H) = 1.22~\AA\ differs from the reference energy by only 0.35~kcal/mol, although the MOB-ML model was not trained on any transition-state like structures.
The highest errors are encountered in the high-energy regime and this region is also associated with the highest Gaussian process variance indicating low confidence in the predictions (compare middle right and right panel of Figure~\ref{figure:figure2}).
The Gaussian process variance reflects the range of structures the MOB-ML model has been trained in and highlights again that we did not include transition-state-like structures in the training. 

\onecolumngrid
\twocolumngrid

\subsection{Transferability across organic chemistry space}
\label{subsec:sec2}

The Chemical Space Project\cite{reymond_chemical_2015} computationally enumerated all possible organic molecules up to a certain number of atoms, resulting in the GDB databases.\cite{blum_970_2009}
In this work, we examine thermalized subsets \cite{cheng_universal_2019} of the GDB13 data set \cite{blum_970_2009} to investigate the transferability of MOB-ML models across organic chemistry space. 
The application of thermalized sets of molecules has the advantage that we can study the transferability of our models for chemical and conformational degrees of freedom at the same time. 
To test the transferability of MOB-ML across chemical space, we train our models on a thermalized set of seven and fewer heavy-atom molecules (also known as QM7b-T \cite{cheng_universal_2019}) and then we test the prediction accuracy on a QM7b-T test set and on a thermalized set of molecules with thirteen heavy atoms (GDB13-T \cite{cheng_universal_2019}; see also Section V in the Supporting Information), as also outlined in our previous work. \cite{cheng_universal_2019, cheng_regression_2019}

\begin{figure}[htbp]
\includegraphics[width=\columnwidth]{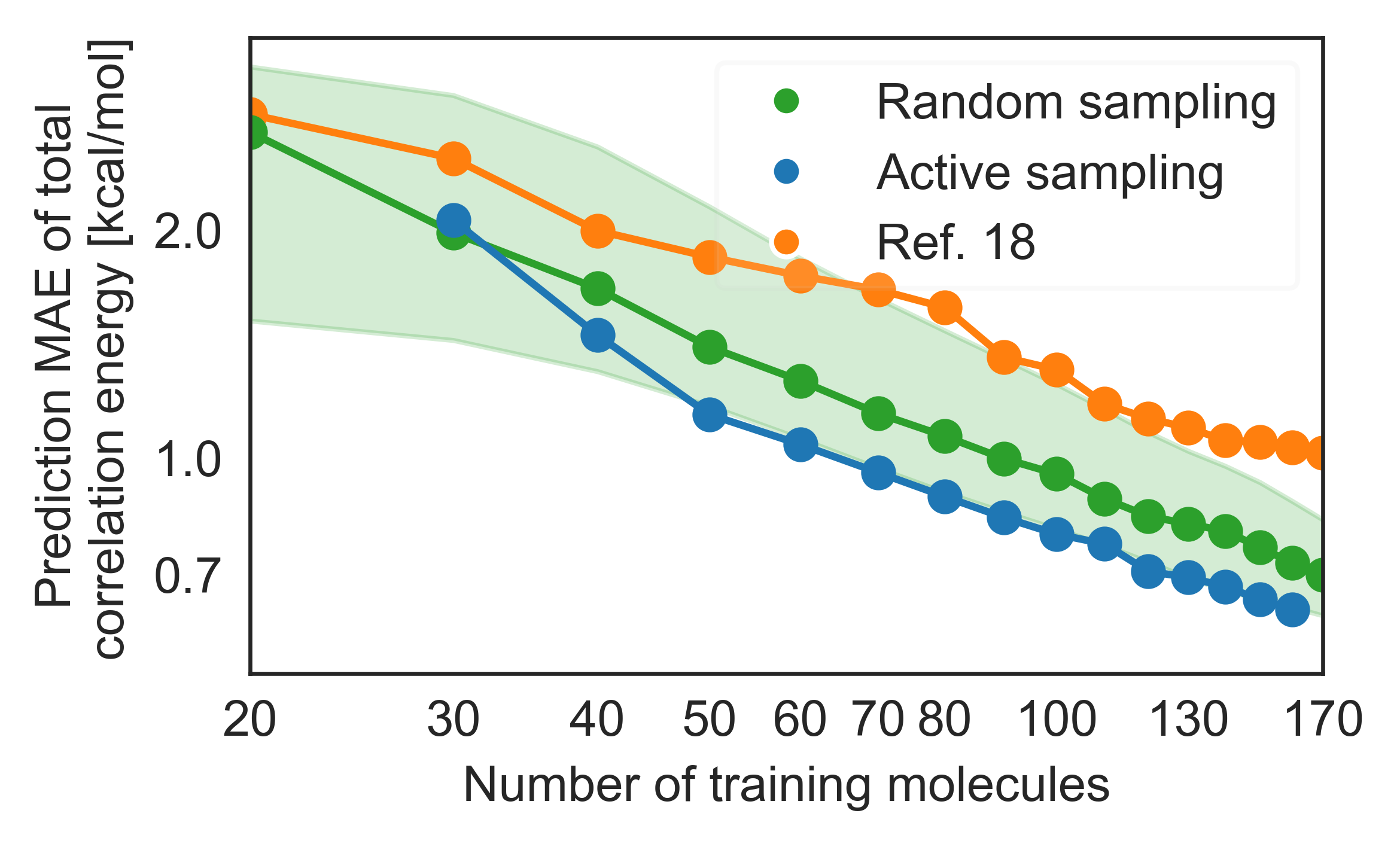}
\caption{
Comparison of the prediction mean absolute errors of total correlation energies for QM7b-T test molecules as a function of the number of QM7b-T molecules chosen for training for different machine learning models: MOB-ML as outlined in Ref.~\onlinecite{cheng_universal_2019} (orange circles), MOB-ML as outlined in this work with random sampling (green circles), and MOB-ML as outlined in this work with active sampling. The green shaded area corresponds to the 90\% confidence interval for the predictions obtained from 50 random samples of the training data.
}
\label{figure:figure3}
\end{figure}

We first investigate the effect of changing the feature vector generation protocol on the QM7b-T$\rightarrow$QM7b-T prediction task (see Figure~\ref{figure:figure3}).
In Ref.~\onlinecite{cheng_universal_2019}, we found that training on about 180 structures is necessary to achieve a model with an MAE below 1~kcal/mol. 
The FHCL method yields an MAE below 1~kcal/mol when training on about 800 structures \cite{christensen_fchl_2020} and the DeepHF method already exhibits an MAE below 1~kcal/mol when training on their smallest chosen training set which consists of 300 structures (MAE=0.79~kcal/mol). \cite{chen_ground_2020} 
The refinements in the current work reduce the number of required training structures to reach chemical accuracy to about 100 structures when sampling randomly.
This number is, however, strongly training set dependent.
We can remove the training-set dependence by switching to an active learning strategy where we can achieve an MAE below 1~kcal/mol reliably with about 70 structures.
In general, the MAE obtained with the active learning strategy is comparable to the smallest MAEs obtained with random sampling strategies.
This has the advantage that a small number of reference data can be generated in a targeted manner.

\begin{figure}[htbp]
\includegraphics[width=\columnwidth]{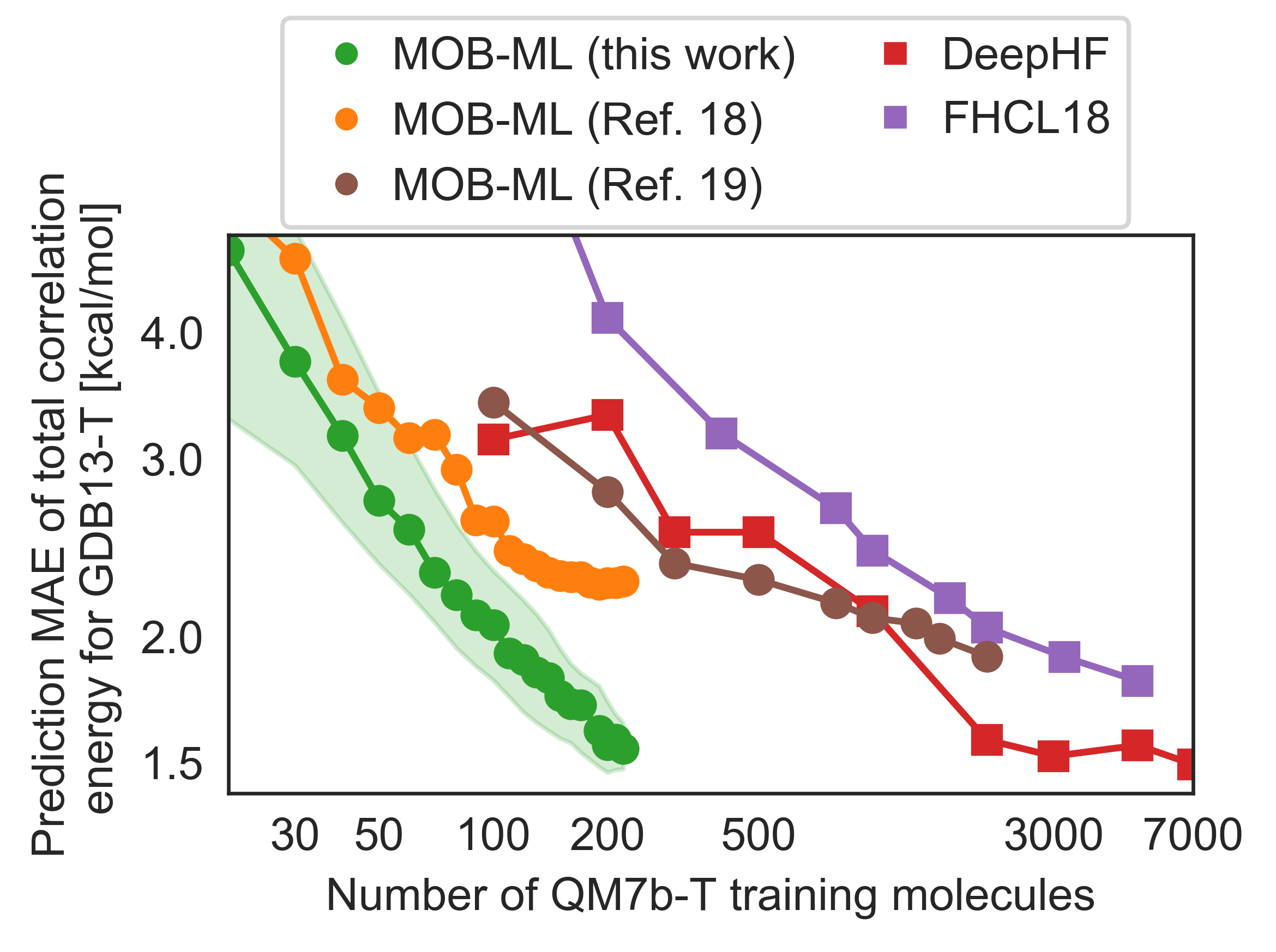}
\caption{
Comparison of the prediction mean absolute errors of total correlation energies for GDB13-T molecules as a function of the number of QM7b-T molecules chosen for model training for different machine learning models: MOB-ML as outlined in this work with random sampling (green circles), 
MOB-ML with a single GPR\cite{cheng_universal_2019} (orange circles), MOB-ML with RCR/GPR \cite{cheng_regression_2019} (brown circles), DeepHF\cite{chen_ground_2020} (red squares), FHCL18\cite{christensen_fchl_2020} (purple squares).
The green shaded area corresponds to the 90\% confidence interval for the predictions obtained from 50 random samples of the training data.
}
\label{figure:figure4}
\end{figure}

In general, our aim is to obtain a machine learning model which reliably predicts broad swathes of chemical space. 
For an ML model to be of practical use, it has to be able to describe out-of-set molecules of different sizes to a similar accuracy when accuracy is measured size-intensively. \cite{SzaboSizeConsistent}
We probe the ability of MOB-ML to describe out-of-set molecules with a different number of electron pairs by applying a model trained on correlation energies for QM7b-T molecules to predict correlation energies for GDB13-T.
We collect the best results published for this transfer test in the literature in Figure~\ref{figure:figure4}.
Our previous best single GPR model achieved an MAE of 2.27~kcal/mol when trained on 220 randomly chosen structures. \cite{cheng_universal_2019} 
The modifications in the current work now yield a single GPR model which achieves an MAE of 1.47--1.62~kcal/mol for GDB13-T when trained on 220 randomly chosen QM7b-T structures.
Strikingly, MOB-ML outperforms machine learning models trained on thousands of molecules like our RCR/GPR model and FHCL18 \cite{christensen_fchl_2020}. The current MOB-ML results are of an accuracy that is similar to the best reported results from DeepHF (an MAE of 1.49 kcal/mol);\cite{chen_ground_2020}
however, MOB-ML only needs to be trained on about 3\% of the molecules in the QM7b data set while DeepHF is trained on 42\% to obtain comparable results (MAE of 1.52~kcal/mol for 3000 training structures). The best reported result for DeepHF (MAE of 1.49 kcal/mol) was obtained by training on 97\% of the molecules of the QM7b data set.
We attribute the excellent transferability of MOB-ML to the fact that it focuses on the prediction of orbital-pair contributions, 
thereby reframing an extrapolation problem into an interpolation problem when training machine learning models on small molecules and testing them on large molecules. 
The pair correlation energies predicted for QM7b-T and for GDB13-T span a very similar range (0 to $-20$~kcal/mol), and they are predicted with a similar Gaussian process variance (see Figure~S5) which we would expect in an interpolation task.
The final errors for GDB13-T are larger than for QM7b-T, because the total correlation energy is size-extensive; however, the size-intensive error per electron pair spans a comparable range for QM7b-T and for GDB13-T (see Figure~S4). 
This presents a significant advantage of MOB-ML over machine learning models which rely on a whole-molecule representation and creates the opportunity to study molecules of a size that are beyond the reach of accurate correlated wave function methods.

\begin{figure}[htbp]
\includegraphics[width=\columnwidth]{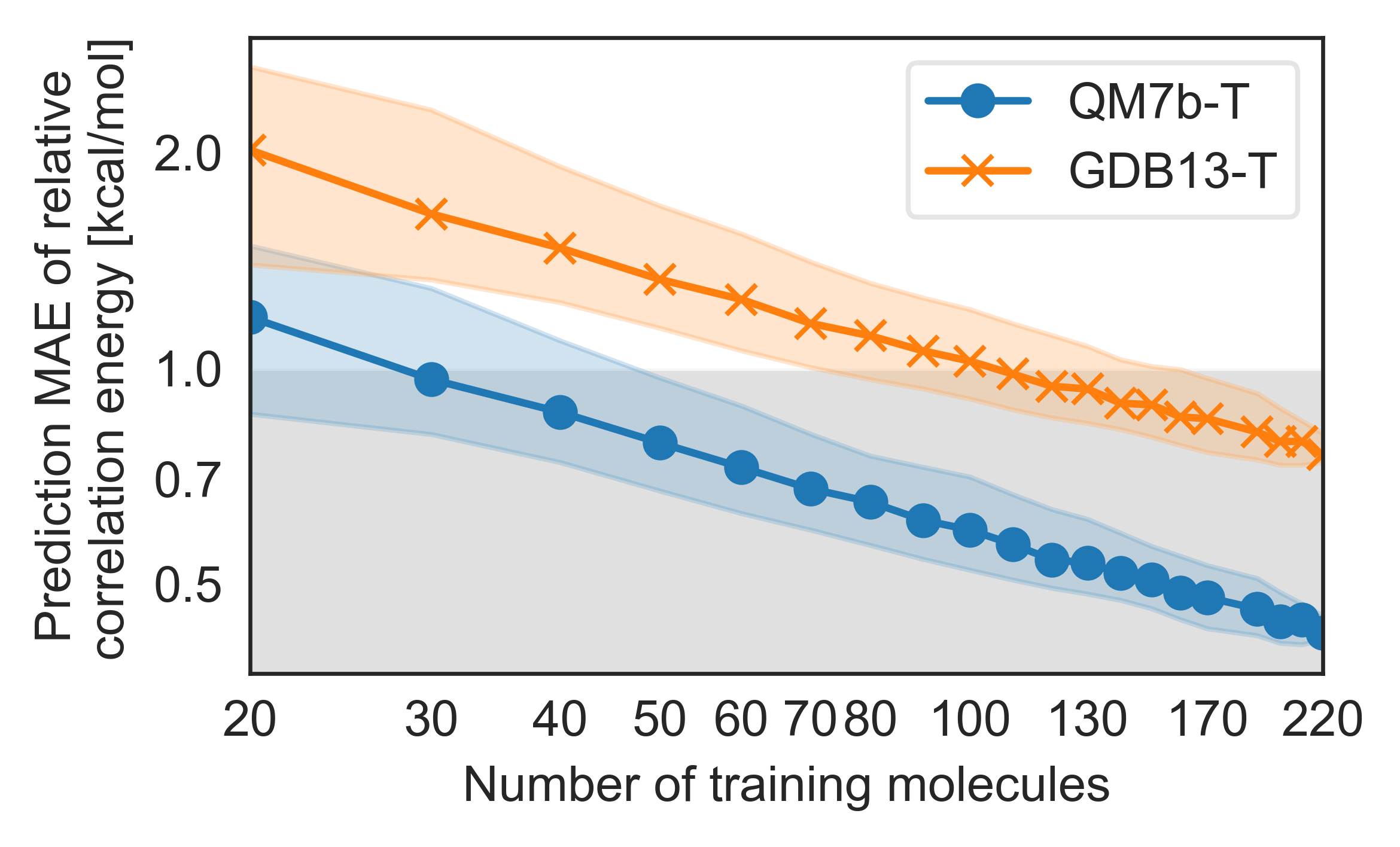}
\caption{
Prediction mean absolute errors for relative correlation energies  as a function of the number of QM7b-T molecules chosen for model training for QM7b-T (blue circles) and for GDB13-T (orange crosses). 
The blue and orange shaded areas correspond to the 90\% confidence interval for the predictions obtained from 50 random samples of the training data.
The gray shaded area corresponds to the region where the error is smaller than chemical accuracy (1~kcal/mol). 
}
\label{figure:figure5}
\end{figure}

Most studies in computational chemistry require accurate relative energies rather than accurate total energies. 
Therefore, we also assess the errors in the relative energies for the sets of conformers for each molecule in the QM7b-T and in the GDB13-T data sets obtained with MOB-ML with respect to the reference energies (see Figure~\ref{figure:figure5}).
We emphasize that MOB-ML is not explicitly trained to predict conformer energies, and we include at most one conformer for each molecule in the training set.
Nevertheless, MOB-ML produces on average chemically accurate relative conformer energies for QM7b-T when trained on correlation energies for only 30 randomly chosen molecules (or 0.4\% of the molecules) in the QM7b set.
We obtain chemically accurate relative energies for the GDB13-T data set when training on about 100 QM7b-T molecules.
The prediction accuracy improves steadily when training on more QM7b-T molecules reaching a mean MAE of 0.43~kcal/mol for the relative energies of the rest of the QM7b-T set and of 0.77~kcal/mol for the GDB13-T set.

\begin{figure}[htbp]
\includegraphics[width=\columnwidth]{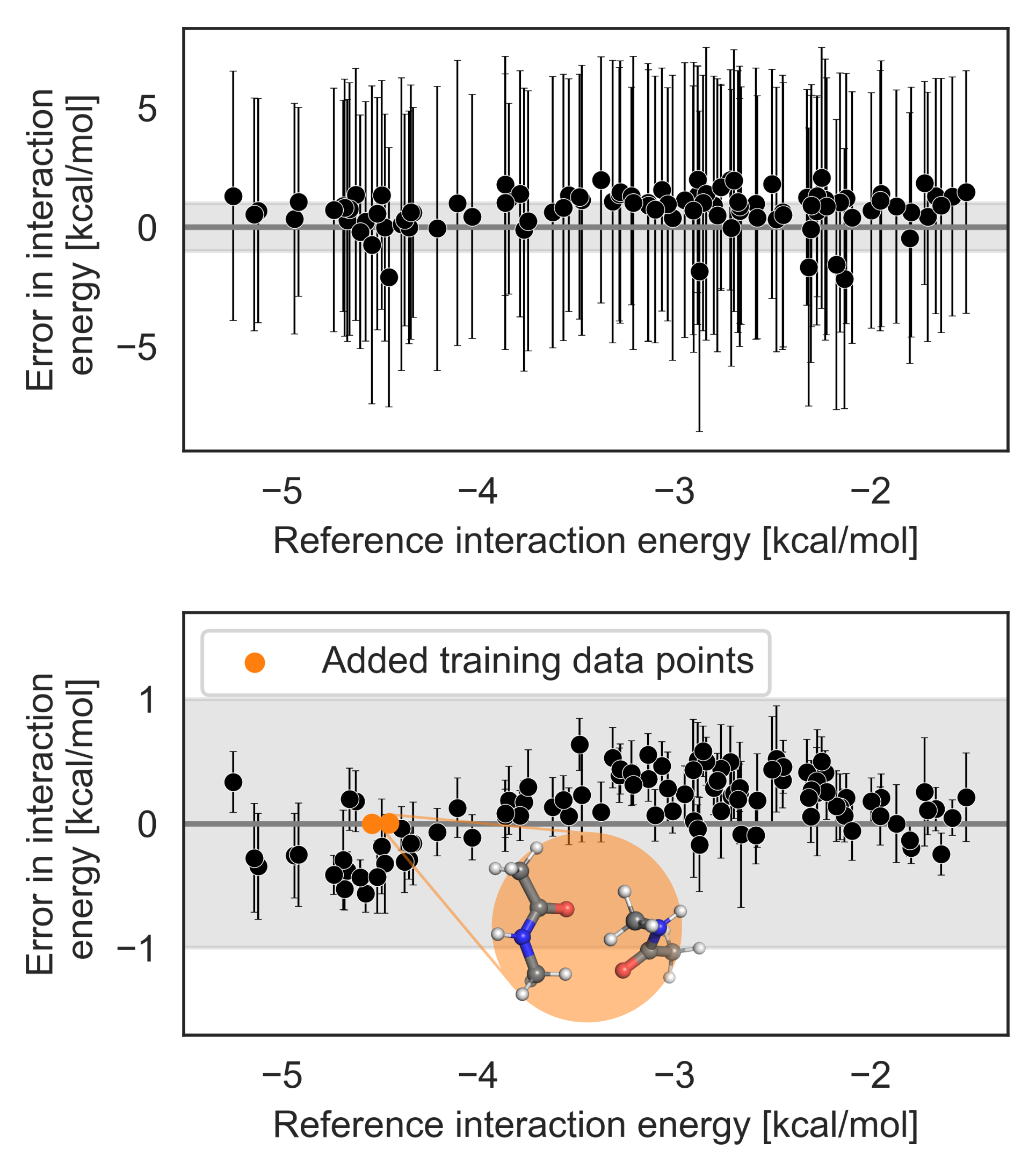}
\caption{
Top panel: Errors in predictions were made with a MOB-ML model trained on 20 randomly selected QM7b-T molecules with FS~3 with respect to reference MP2/cc-pVTZ interaction energies for the BBI data set.
Bottom panel:
Errors in predictions were made with a MOB-ML model trained on 20 randomly selected QM7b-T molecules and augmented with the 2 BBI data points with the largest variance (orange circles) with respect to reference MP2/cc-pVTZ interaction energies.
The bar attached to each prediction error indicates the associated Gaussian process variance.
The gray shaded area corresponds to the region where the error is smaller than chemical accuracy (1~kcal/mol). 
}
\label{figure:figure6}
\end{figure}

We now present the first reported test of MOB-ML for non-covalent interactions in large molecules. 
To this end, we examine the backbone-backbone interaction (BBI) data set \cite{burns_biofragment_2017} which was designed to benchmark methods for the prediction of interaction energies encountered within protein fragments.
Using the implementation of MOB-ML described here and using only 20 randomly selected QM7b-T molecules for training, the method achieves a mean absolute error of 0.98~kcal/mol for the BBI data set (see Figure~\ref{figure:figure6}).
However, these predictions are uncertain as indicated by the large Gaussian process variances associated with these data points which strongly suggested that we are now, as expected, in an extrapolative regime. 
We further improve the predictive capability of MOB-ML by augmenting the MOB-ML model with  data from the BBI set. 
Specifically, we can draw on an active learning strategy and consecutively include data points until all uncertainties are below 1~kcal/mol which in this case corresponds to only two data points. 
This reduces the MAE to 0.28~kcal/mol for the remaining 98 data points in the BBI set.
Including more reference data points would further improve the performance for this specific data set. However, this is not the focus of this work. 
Instead, we simply emphasize that MOB-ML is a clearly extensible strategy to accurately predict energies for large molecules and non-covalent intermolecular interactions while providing a useful estimation of confidence.

\subsection{Transition-metal complexes} 
\label{subsec:sec3}

\begin{figure}[h]
\includegraphics[width=\columnwidth]{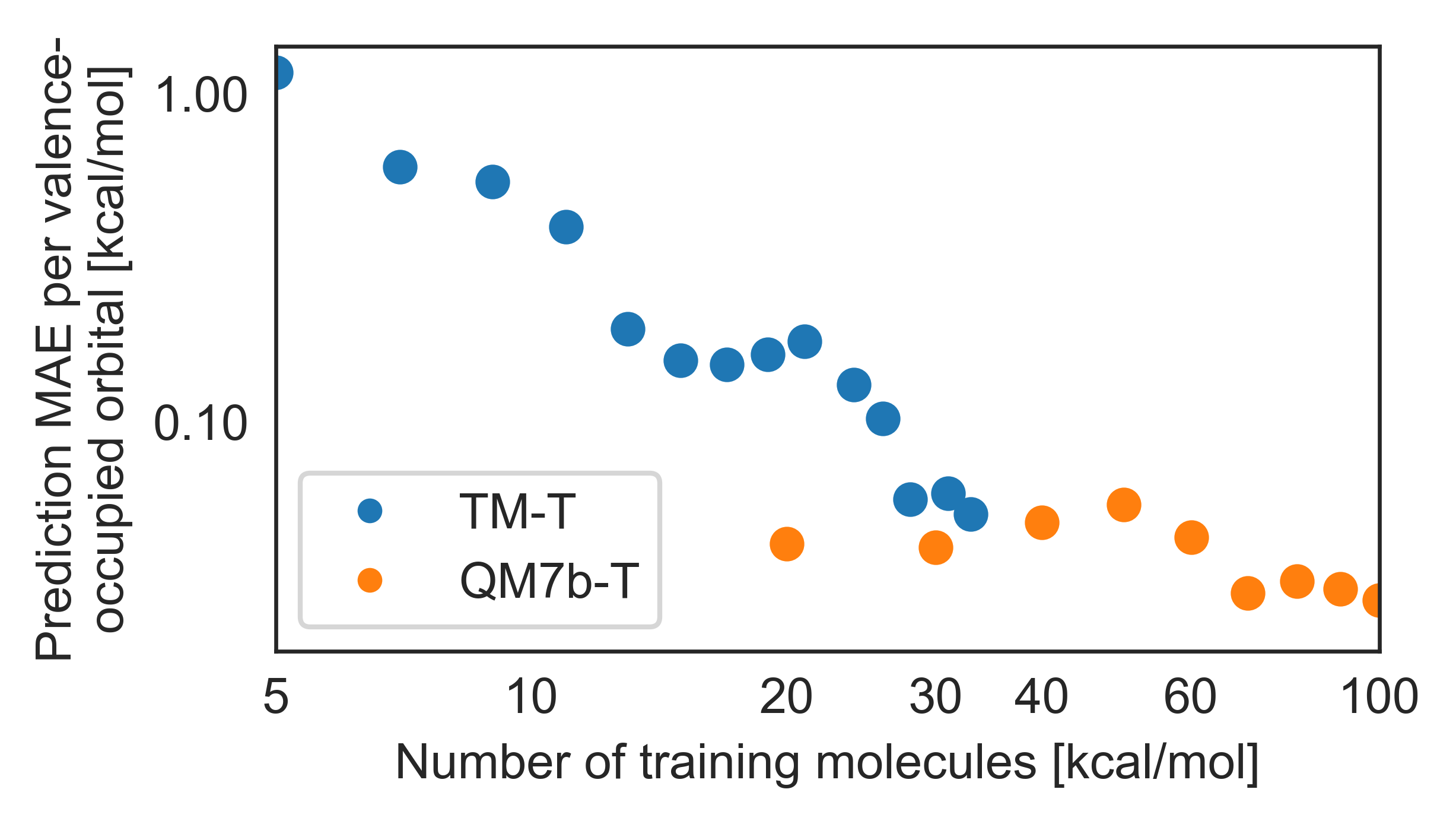}
\caption{
 Learning curve for the prediction of MP2 correlation energies per valence-occupied orbital for transition metal complexes (TM-T) and for QM7b-T as a function of the number of structures the MOB-ML model was trained on.
}
\label{figure:figure7}
\end{figure}

We finally present the first application of MOB-ML to transition-metal complexes. 
To this end, we train a MOB-ML model on a thermalized subset of mononuclear, octahedral transition-metal complexes  introduced by Kulik and co-workers \cite{nandy_strategies_2018} which we denote as TM-T.
The chosen closed-shell transition-metal complexes feature different transition metals (Fe, Co, Ni) and ligands.
The ligands span the spectrochemical series from weak-field (e.g., thiocyanate) over to strong-field (e.g., carbonyl) ligands. 
We see in Figure~\ref{figure:figure7} that the learning behaviour between TM-T and QM7b-T is similar when the error is measured per valence-occupied orbital.
These results demonstrate that MOB-ML formalism can be straightforwardly applied outside of the organic chemistry universe without additional modifications.
It is particularly notable that the learning efficiency for TM-T is comparable to that for QM7b-T, as seen in the relatively simple organic molecules in QM7b-T (Fig.~\ref{figure:figure7}). 
We note that whereas MP2 theory is not expected to be fully quantitative for transition metal complexes, \cite{weymuth_new_2014, husch_calculation_2018}
it provides a demonstration of the learning efficiency of MOB-ML for transition-metal complexes in the current example;
and as previously demonstrated, MOB-ML learns other correlated wave function methods with similar efficiency.\cite{welborn_transferability_2018, cheng_universal_2019}

\section{Conclusions}

Molecular-orbital-based machine learning (MOB-ML) provides a general framework to learn correlation energies at the cost of  molecular orbital generation.
In this work, we  demonstrate that preservation of physical symmetries and constraints leads to machine-learning methods with greater learning efficiency and transferability. 
Exploiting physical principles like size consistency and energy invariances  not only leads to a conceptually more satisfying method, but it also leads to substantial improvements in prediction errors for different data sets covering total and relative energies for thermally accessible organic and transition-metal containing molecules, non-covalent interactions, and transition-state energies. 
With the modifications presented in the current work, 
MOB-ML is shown to be highly data efficient, which is important due to the high computational cost of generating  reference correlation energies.
Only 1\% of the QM7b-T data set (containing organic molecules with seven and fewer heavy atoms) needs to be drawn on to train a MOB-ML model which produces on average chemically accurate total energies for the remaining 99\% of the data set.
Without ever being trained to predict relative energies, MOB-ML provides chemically accurate relative energies for QM7b-T when training on only 0.4\% of the QM7b-T molecules.
Furthermore, we have demonstrated that MOB-ML is not restricted to the organic chemistry space and that we are able to apply our framework out-of-the box to describe a diverse set transition-metal complexes when training on correlation energies for tens of molecules.

Beyond data efficiency, MOB-ML models are are shown to be very transferable across chemical space. 
We demonstrate this transferability by training a MOB-ML model on QM7b-T and predicting energies for a set of molecules with thirteen heavy atoms (GDB13-T). 
We obtain the best result for GDB13-T reported to date despite only training on 3\% of QM7b-T. 
The successful transferability of MOB-ML  is shown to result from its recasting of a typical extrapolation task (i.e., larger molecules) into an interpolation task (i.e., by predicting on the basis of size-intensive orbital-pair contributions).
Even when MOB-ML enters an extrapolative regime as identified by a large Gaussian process variance, accurate results can be obtained; for example, we predict the transition-state energy for the proton transfer in malonaldehyde and interaction energies in the protein backbone-backbone interaction data set to chemical accuracy without training on transition-state-like data or non-covalent interactions, respectively.
In this case, the uncertainty estimates also offer a clear avenue for active learning strategies which can further improve the model performance.
Active learning offers an attractive way to reduce the number of expensive reference calculations further by picking the most informative molecules to be included in the training set.  
This provides a general recipe how to evolve a MOB-ML model to describe new regions of chemical space with minimal effort. 

Future work will focus on the expansion of MOB-ML to cover more of chemical space. 
Specifically, particular areas of focus include open-shell systems and electronically excited states. 
Physical insight from exact conditions in electronic structure theory \cite{bartlett_power_2017} will continue to guide the development of the method, with the aim of providing a machine-learning approach for energies and properties of arbitrary molecules with controlled error. 

\begin{acknowledgments}
This work is supported in part by the U.S. Army Research Laboratory (W911NF-12-2-0023), the U.S. Department of Energy (DE-SC0019390), the Caltech DeLogi Fund, and the Camille and Henry Dreyfus Foundation (Award ML-20-196). 
T.H. acknowledges funding through an Early Post-Doc Mobility Fellowship  by  the  Swiss  National  Science Foundation (Award P2EZP2\_184234). 
S.J.R.L. thanks the Molecular Software Sciences Institute (MolSSI) for a MolSSI investment fellowship. 
Computational resources were provided by the National Energy Research Scientific
Computing Center (NERSC), a DOE Office of Science User Facility
supported by the DOE Office of Science under contract DE-AC02-05CH11231.
\end{acknowledgments}

\section*{Supporting Information}
Details on feature generation for all data sets used in this work, definition of error metrics, expanded results for the alkane transferability test, expanded results for the transferability within the organic chemistry space.
Features and labels for all data sets used in this work.

\section*{Data Availability Statement}
The data that supports the findings of this study are available within the article and its supplementary material.
Additional data that support the findings of this study are openly available in Caltech Data Repository.\cite{caltech_data}

\bibliography{main}

\begin{thebibliography}{64}%
\makeatletter
\providecommand \@ifxundefined [1]{%
 \@ifx{#1\undefined}
}%
\providecommand \@ifnum [1]{%
 \ifnum #1\expandafter \@firstoftwo
 \else \expandafter \@secondoftwo
 \fi
}%
\providecommand \@ifx [1]{%
 \ifx #1\expandafter \@firstoftwo
 \else \expandafter \@secondoftwo
 \fi
}%
\providecommand \natexlab [1]{#1}%
\providecommand \enquote  [1]{``#1''}%
\providecommand \bibnamefont  [1]{#1}%
\providecommand \bibfnamefont [1]{#1}%
\providecommand \citenamefont [1]{#1}%
\providecommand \href@noop [0]{\@secondoftwo}%
\providecommand \href [0]{\begingroup \@sanitize@url \@href}%
\providecommand \@href[1]{\@@startlink{#1}\@@href}%
\providecommand \@@href[1]{\endgroup#1\@@endlink}%
\providecommand \@sanitize@url [0]{\catcode `\\12\catcode `\$12\catcode
  `\&12\catcode `\#12\catcode `\^12\catcode `\_12\catcode `\%12\relax}%
\providecommand \@@startlink[1]{}%
\providecommand \@@endlink[0]{}%
\providecommand \url  [0]{\begingroup\@sanitize@url \@url }%
\providecommand \@url [1]{\endgroup\@href {#1}{\urlprefix }}%
\providecommand \urlprefix  [0]{URL }%
\providecommand \Eprint [0]{\href }%
\providecommand \doibase [0]{http://dx.doi.org/}%
\providecommand \selectlanguage [0]{\@gobble}%
\providecommand \bibinfo  [0]{\@secondoftwo}%
\providecommand \bibfield  [0]{\@secondoftwo}%
\providecommand \translation [1]{[#1]}%
\providecommand \BibitemOpen [0]{}%
\providecommand \bibitemStop [0]{}%
\providecommand \bibitemNoStop [0]{.\EOS\space}%
\providecommand \EOS [0]{\spacefactor3000\relax}%
\providecommand \BibitemShut  [1]{\csname bibitem#1\endcsname}%
\let\auto@bib@innerbib\@empty
\bibitem [{\citenamefont {Bartók}\ \emph {et~al.}(2010)\citenamefont
  {Bartók}, \citenamefont {Payne}, \citenamefont {Kondor},\ and\ \citenamefont
  {Csányi}}]{bartok_gaussian_2010}%
  \BibitemOpen
  \bibfield  {author} {\bibinfo {author} {\bibfnamefont {A.~P.}\ \bibnamefont
  {Bartók}}, \bibinfo {author} {\bibfnamefont {M.~C.}\ \bibnamefont {Payne}},
  \bibinfo {author} {\bibfnamefont {R.}~\bibnamefont {Kondor}}, \ and\ \bibinfo
  {author} {\bibfnamefont {G.}~\bibnamefont {Csányi}},\ }\bibfield  {title}
  {\enquote {\bibinfo {title} {Gaussian {Approximation} {Potentials}: {The}
  {Accuracy} of {Quantum} {Mechanics}, without the {Electrons}},}\ }\href@noop
  {} {\bibfield  {journal} {\bibinfo  {journal} {Phys. Rev. Lett.}\ }\textbf
  {\bibinfo {volume} {104}},\ \bibinfo {pages} {136403} (\bibinfo {year}
  {2010})}\BibitemShut {NoStop}%
\bibitem [{\citenamefont {Rupp}\ \emph {et~al.}(2012)\citenamefont {Rupp},
  \citenamefont {Tkatchenko}, \citenamefont {Müller},\ and\ \citenamefont {von
  Lilienfeld}}]{rupp_fast_2012}%
  \BibitemOpen
  \bibfield  {author} {\bibinfo {author} {\bibfnamefont {M.}~\bibnamefont
  {Rupp}}, \bibinfo {author} {\bibfnamefont {A.}~\bibnamefont {Tkatchenko}},
  \bibinfo {author} {\bibfnamefont {K.-R.}\ \bibnamefont {Müller}}, \ and\
  \bibinfo {author} {\bibfnamefont {O.~A.}\ \bibnamefont {von Lilienfeld}},\
  }\bibfield  {title} {\enquote {\bibinfo {title} {Fast and {Accurate}
  {Modeling} of {Molecular} {Atomization} {Energies} with {Machine}
  {Learning}},}\ }\href@noop {} {\bibfield  {journal} {\bibinfo  {journal}
  {Phys. Rev. Lett.}\ }\textbf {\bibinfo {volume} {108}},\ \bibinfo {pages}
  {058301} (\bibinfo {year} {2012})}\BibitemShut {NoStop}%
\bibitem [{\citenamefont {Hansen}\ \emph {et~al.}(2013)\citenamefont {Hansen},
  \citenamefont {Montavon}, \citenamefont {Biegler}, \citenamefont {Fazli},
  \citenamefont {Rupp}, \citenamefont {Scheffler}, \citenamefont {von
  Lilienfeld}, \citenamefont {Tkatchenko},\ and\ \citenamefont
  {Müller}}]{hansen_assessment_2013}%
  \BibitemOpen
  \bibfield  {author} {\bibinfo {author} {\bibfnamefont {K.}~\bibnamefont
  {Hansen}}, \bibinfo {author} {\bibfnamefont {G.}~\bibnamefont {Montavon}},
  \bibinfo {author} {\bibfnamefont {F.}~\bibnamefont {Biegler}}, \bibinfo
  {author} {\bibfnamefont {S.}~\bibnamefont {Fazli}}, \bibinfo {author}
  {\bibfnamefont {M.}~\bibnamefont {Rupp}}, \bibinfo {author} {\bibfnamefont
  {M.}~\bibnamefont {Scheffler}}, \bibinfo {author} {\bibfnamefont {O.~A.}\
  \bibnamefont {von Lilienfeld}}, \bibinfo {author} {\bibfnamefont
  {A.}~\bibnamefont {Tkatchenko}}, \ and\ \bibinfo {author} {\bibfnamefont
  {K.-R.}\ \bibnamefont {Müller}},\ }\bibfield  {title} {\enquote {\bibinfo
  {title} {Assessment and validation of machine learning methods for predicting
  molecular atomization energies},}\ }\href@noop {} {\bibfield  {journal}
  {\bibinfo  {journal} {J. Chem. Theory Comput.}\ }\textbf {\bibinfo {volume}
  {9}},\ \bibinfo {pages} {3404} (\bibinfo {year} {2013})}\BibitemShut
  {NoStop}%
\bibitem [{\citenamefont {Ramakrishnan}\ \emph {et~al.}(2015)\citenamefont
  {Ramakrishnan}, \citenamefont {Dral}, \citenamefont {Rupp},\ and\
  \citenamefont {von Lilienfeld}}]{ramakrishnan_big_2015}%
  \BibitemOpen
  \bibfield  {author} {\bibinfo {author} {\bibfnamefont {R.}~\bibnamefont
  {Ramakrishnan}}, \bibinfo {author} {\bibfnamefont {P.~O.}\ \bibnamefont
  {Dral}}, \bibinfo {author} {\bibfnamefont {M.}~\bibnamefont {Rupp}}, \ and\
  \bibinfo {author} {\bibfnamefont {O.~A.}\ \bibnamefont {von Lilienfeld}},\
  }\bibfield  {title} {\enquote {\bibinfo {title} {Big {Data} {Meets} {Quantum}
  {Chemistry} {Approximations}: {The} $\delta$-{Machine} {Learning}
  {Approach}},}\ }\href@noop {} {\bibfield  {journal} {\bibinfo  {journal} {J.
  Chem. Theory Comput.}\ }\textbf {\bibinfo {volume} {11}},\ \bibinfo {pages}
  {2087--2096} (\bibinfo {year} {2015})}\BibitemShut {NoStop}%
\bibitem [{\citenamefont {Behler}(2016)}]{behler_perspective_2016}%
  \BibitemOpen
  \bibfield  {author} {\bibinfo {author} {\bibfnamefont {J.}~\bibnamefont
  {Behler}},\ }\bibfield  {title} {\enquote {\bibinfo {title} {Perspective:
  {Machine} learning potentials for atomistic simulations},}\ }\href@noop {}
  {\bibfield  {journal} {\bibinfo  {journal} {J. Chem. Phys.}\ }\textbf
  {\bibinfo {volume} {145}},\ \bibinfo {pages} {170901} (\bibinfo {year}
  {2016})}\BibitemShut {NoStop}%
\bibitem [{\citenamefont {Brockherde}\ \emph {et~al.}(2017)\citenamefont
  {Brockherde}, \citenamefont {Vogt}, \citenamefont {Li}, \citenamefont
  {Tuckerman}, \citenamefont {Burke},\ and\ \citenamefont
  {Müller}}]{brockherde_bypassing_2017}%
  \BibitemOpen
  \bibfield  {author} {\bibinfo {author} {\bibfnamefont {F.}~\bibnamefont
  {Brockherde}}, \bibinfo {author} {\bibfnamefont {L.}~\bibnamefont {Vogt}},
  \bibinfo {author} {\bibfnamefont {L.}~\bibnamefont {Li}}, \bibinfo {author}
  {\bibfnamefont {M.~E.}\ \bibnamefont {Tuckerman}}, \bibinfo {author}
  {\bibfnamefont {K.}~\bibnamefont {Burke}}, \ and\ \bibinfo {author}
  {\bibfnamefont {K.-R.}\ \bibnamefont {Müller}},\ }\bibfield  {title}
  {\enquote {\bibinfo {title} {Bypassing the {Kohn}-{Sham} equations with
  machine learning},}\ }\href@noop {} {\bibfield  {journal} {\bibinfo
  {journal} {Nat. Comm.}\ }\textbf {\bibinfo {volume} {8}},\ \bibinfo {pages}
  {872} (\bibinfo {year} {2017})}\BibitemShut {NoStop}%
\bibitem [{\citenamefont {Schütt}\ \emph
  {et~al.}(2017{\natexlab{a}})\citenamefont {Schütt}, \citenamefont
  {Kindermans}, \citenamefont {Sauceda}, \citenamefont {Chmiela}, \citenamefont
  {Tkatchenko},\ and\ \citenamefont {Müller}}]{schutt_schnet_2017}%
  \BibitemOpen
  \bibfield  {author} {\bibinfo {author} {\bibfnamefont {K.}~\bibnamefont
  {Schütt}}, \bibinfo {author} {\bibfnamefont {P.-J.}\ \bibnamefont
  {Kindermans}}, \bibinfo {author} {\bibfnamefont {H.~E.}\ \bibnamefont
  {Sauceda}}, \bibinfo {author} {\bibfnamefont {S.}~\bibnamefont {Chmiela}},
  \bibinfo {author} {\bibfnamefont {A.}~\bibnamefont {Tkatchenko}}, \ and\
  \bibinfo {author} {\bibfnamefont {K.-R.}\ \bibnamefont {Müller}},\
  }\bibfield  {title} {\enquote {\bibinfo {title} {{SchNet}: {A}
  continuous-filter convolutional neural network for modeling quantum
  interactions},}\ }in\ \href@noop {} {\emph {\bibinfo {booktitle} {Advances in
  {Neural} {Information} {Processing} {Systems} 30}}},\ \bibinfo {editor}
  {edited by\ \bibinfo {editor} {\bibfnamefont {I.}~\bibnamefont {Guyon}},
  \bibinfo {editor} {\bibfnamefont {U.~V.}\ \bibnamefont {Luxburg}}, \bibinfo
  {editor} {\bibfnamefont {S.}~\bibnamefont {Bengio}}, \bibinfo {editor}
  {\bibfnamefont {H.}~\bibnamefont {Wallach}}, \bibinfo {editor} {\bibfnamefont
  {R.}~\bibnamefont {Fergus}}, \bibinfo {editor} {\bibfnamefont
  {S.}~\bibnamefont {Vishwanathan}}, \ and\ \bibinfo {editor} {\bibfnamefont
  {R.}~\bibnamefont {Garnett}}}\ (\bibinfo  {publisher} {Curran Associates,
  Inc.},\ \bibinfo {year} {2017})\ pp.\ \bibinfo {pages}
  {991--1001}\BibitemShut {NoStop}%
\bibitem [{\citenamefont {Schütt}\ \emph
  {et~al.}(2017{\natexlab{b}})\citenamefont {Schütt}, \citenamefont
  {Arbabzadah}, \citenamefont {Chmiela}, \citenamefont {Müller},\ and\
  \citenamefont {Tkatchenko}}]{schutt_quantum-chemical_2017}%
  \BibitemOpen
  \bibfield  {author} {\bibinfo {author} {\bibfnamefont {K.~T.}\ \bibnamefont
  {Schütt}}, \bibinfo {author} {\bibfnamefont {F.}~\bibnamefont {Arbabzadah}},
  \bibinfo {author} {\bibfnamefont {S.}~\bibnamefont {Chmiela}}, \bibinfo
  {author} {\bibfnamefont {K.-R.}\ \bibnamefont {Müller}}, \ and\ \bibinfo
  {author} {\bibfnamefont {A.}~\bibnamefont {Tkatchenko}},\ }\bibfield  {title}
  {\enquote {\bibinfo {title} {Quantum-chemical insights from deep tensor
  neural networks},}\ }\href@noop {} {\bibfield  {journal} {\bibinfo  {journal}
  {Nat. Commun.}\ }\textbf {\bibinfo {volume} {8}},\ \bibinfo {pages} {13890}
  (\bibinfo {year} {2017}{\natexlab{b}})}\BibitemShut {NoStop}%
\bibitem [{\citenamefont {Smith}, \citenamefont {Isayev},\ and\ \citenamefont
  {Roitberg}(2017)}]{smith_ani-1_2017}%
  \BibitemOpen
  \bibfield  {author} {\bibinfo {author} {\bibfnamefont {J.~S.}\ \bibnamefont
  {Smith}}, \bibinfo {author} {\bibfnamefont {O.}~\bibnamefont {Isayev}}, \
  and\ \bibinfo {author} {\bibfnamefont {A.~E.}\ \bibnamefont {Roitberg}},\
  }\bibfield  {title} {\enquote {\bibinfo {title} {{ANI}-1: an extensible
  neural network potential with {DFT} accuracy at force field computational
  cost},}\ }\href@noop {} {\bibfield  {journal} {\bibinfo  {journal} {Chem.
  Sci.}\ }\textbf {\bibinfo {volume} {8}},\ \bibinfo {pages} {3192--3203}
  (\bibinfo {year} {2017})}\BibitemShut {NoStop}%
\bibitem [{\citenamefont {McGibbon}\ \emph {et~al.}(2017)\citenamefont
  {McGibbon}, \citenamefont {Taube}, \citenamefont {Donchev}, \citenamefont
  {Siva}, \citenamefont {Hern{'a}ndez}, \citenamefont {Hargus}, \citenamefont
  {Law}, \citenamefont {Klepeis},\ and\ \citenamefont
  {Shaw}}]{mcgibbon_improving_2017}%
  \BibitemOpen
  \bibfield  {author} {\bibinfo {author} {\bibfnamefont {R.~T.}\ \bibnamefont
  {McGibbon}}, \bibinfo {author} {\bibfnamefont {A.~G.}\ \bibnamefont {Taube}},
  \bibinfo {author} {\bibfnamefont {A.~G.}\ \bibnamefont {Donchev}}, \bibinfo
  {author} {\bibfnamefont {K.}~\bibnamefont {Siva}}, \bibinfo {author}
  {\bibfnamefont {F.}~\bibnamefont {Hern{'a}ndez}}, \bibinfo {author}
  {\bibfnamefont {C.}~\bibnamefont {Hargus}}, \bibinfo {author} {\bibfnamefont
  {K.-H.}\ \bibnamefont {Law}}, \bibinfo {author} {\bibfnamefont {J.~L.}\
  \bibnamefont {Klepeis}}, \ and\ \bibinfo {author} {\bibfnamefont {D.~E.}\
  \bibnamefont {Shaw}},\ }\bibfield  {title} {\enquote {\bibinfo {title}
  {Improving the accuracy of m{\o}ller-plesset perturbation theory with neural
  networks},}\ }\href@noop {} {\bibfield  {journal} {\bibinfo  {journal} {J.
  Chem. Phys.}\ }\textbf {\bibinfo {volume} {147}},\ \bibinfo {pages} {161725}
  (\bibinfo {year} {2017})}\BibitemShut {NoStop}%
\bibitem [{\citenamefont {Collins}\ \emph {et~al.}(2018)\citenamefont
  {Collins}, \citenamefont {Gordon}, \citenamefont {von Lilienfeld},\ and\
  \citenamefont {Yaron}}]{collins_constant_2018}%
  \BibitemOpen
  \bibfield  {author} {\bibinfo {author} {\bibfnamefont {C.~R.}\ \bibnamefont
  {Collins}}, \bibinfo {author} {\bibfnamefont {G.~J.}\ \bibnamefont {Gordon}},
  \bibinfo {author} {\bibfnamefont {O.~A.}\ \bibnamefont {von Lilienfeld}}, \
  and\ \bibinfo {author} {\bibfnamefont {D.~J.}\ \bibnamefont {Yaron}},\
  }\bibfield  {title} {\enquote {\bibinfo {title} {Constant size descriptors
  for accurate machine learning models of molecular properties},}\ }\href@noop
  {} {\bibfield  {journal} {\bibinfo  {journal} {J. Chem. Phys.}\ }\textbf
  {\bibinfo {volume} {148}},\ \bibinfo {pages} {241718} (\bibinfo {year}
  {2018})}\BibitemShut {NoStop}%
\bibitem [{\citenamefont {Fujikake}\ \emph {et~al.}(2018)\citenamefont
  {Fujikake}, \citenamefont {Deringer}, \citenamefont {Lee}, \citenamefont
  {Krynski}, \citenamefont {Elliott},\ and\ \citenamefont
  {Csányi}}]{fujikake_gaussian_2018}%
  \BibitemOpen
  \bibfield  {author} {\bibinfo {author} {\bibfnamefont {S.}~\bibnamefont
  {Fujikake}}, \bibinfo {author} {\bibfnamefont {V.~L.}\ \bibnamefont
  {Deringer}}, \bibinfo {author} {\bibfnamefont {T.~H.}\ \bibnamefont {Lee}},
  \bibinfo {author} {\bibfnamefont {M.}~\bibnamefont {Krynski}}, \bibinfo
  {author} {\bibfnamefont {S.~R.}\ \bibnamefont {Elliott}}, \ and\ \bibinfo
  {author} {\bibfnamefont {G.}~\bibnamefont {Csányi}},\ }\bibfield  {title}
  {\enquote {\bibinfo {title} {Gaussian approximation potential modeling of
  lithium intercalation in carbon nanostructures},}\ }\href@noop {} {\bibfield
  {journal} {\bibinfo  {journal} {J. Chem. Phys.}\ }\textbf {\bibinfo {volume}
  {148}},\ \bibinfo {pages} {241714} (\bibinfo {year} {2018})}\BibitemShut
  {NoStop}%
\bibitem [{\citenamefont {Lubbers}, \citenamefont {Smith},\ and\ \citenamefont
  {Barros}(2018)}]{lubbers_hierarchical_2018}%
  \BibitemOpen
  \bibfield  {author} {\bibinfo {author} {\bibfnamefont {N.}~\bibnamefont
  {Lubbers}}, \bibinfo {author} {\bibfnamefont {J.~S.}\ \bibnamefont {Smith}},
  \ and\ \bibinfo {author} {\bibfnamefont {K.}~\bibnamefont {Barros}},\
  }\bibfield  {title} {\enquote {\bibinfo {title} {Hierarchical modeling of
  molecular energies using a deep neural network},}\ }\href@noop {} {\bibfield
  {journal} {\bibinfo  {journal} {J. Chem. Phys.}\ }\textbf {\bibinfo {volume}
  {148}},\ \bibinfo {pages} {241715} (\bibinfo {year} {2018})}\BibitemShut
  {NoStop}%
\bibitem [{\citenamefont {Nguyen}\ \emph {et~al.}(2018)\citenamefont {Nguyen},
  \citenamefont {Székely}, \citenamefont {Imbalzano}, \citenamefont {Behler},
  \citenamefont {Csányi}, \citenamefont {Ceriotti}, \citenamefont {Götz},\
  and\ \citenamefont {Paesani}}]{nguyen_comparison_2018}%
  \BibitemOpen
  \bibfield  {author} {\bibinfo {author} {\bibfnamefont {T.~T.}\ \bibnamefont
  {Nguyen}}, \bibinfo {author} {\bibfnamefont {E.}~\bibnamefont {Székely}},
  \bibinfo {author} {\bibfnamefont {G.}~\bibnamefont {Imbalzano}}, \bibinfo
  {author} {\bibfnamefont {{\"o}.}~\bibnamefont {Behler}}, \bibinfo {author}
  {\bibfnamefont {G.}~\bibnamefont {Csányi}}, \bibinfo {author} {\bibfnamefont
  {M.}~\bibnamefont {Ceriotti}}, \bibinfo {author} {\bibfnamefont {A.~W.}\
  \bibnamefont {Götz}}, \ and\ \bibinfo {author} {\bibfnamefont
  {F.}~\bibnamefont {Paesani}},\ }\bibfield  {title} {\enquote {\bibinfo
  {title} {Comparison of permutationally invariant polynomials, neural
  networks, and {Gaussian} approximation potentials in representing water
  interactions through many-body expansions},}\ }\href@noop {} {\bibfield
  {journal} {\bibinfo  {journal} {J. Chem. Phys.}\ }\textbf {\bibinfo {volume}
  {148}},\ \bibinfo {pages} {241725} (\bibinfo {year} {2018})}\BibitemShut
  {NoStop}%
\bibitem [{\citenamefont {Welborn}, \citenamefont {Cheng},\ and\ \citenamefont
  {Miller~III}(2018)}]{welborn_transferability_2018}%
  \BibitemOpen
  \bibfield  {author} {\bibinfo {author} {\bibfnamefont {M.}~\bibnamefont
  {Welborn}}, \bibinfo {author} {\bibfnamefont {L.}~\bibnamefont {Cheng}}, \
  and\ \bibinfo {author} {\bibfnamefont {T.~F.}\ \bibnamefont {Miller~III}},\
  }\bibfield  {title} {\enquote {\bibinfo {title} {Transferability in {Machine}
  {Learning} for {Electronic} {Structure} via the {Molecular} {Orbital}
  {Basis}},}\ }\href@noop {} {\bibfield  {journal} {\bibinfo  {journal} {J.
  Chem. Theory Comput.}\ }\textbf {\bibinfo {volume} {14}},\ \bibinfo {pages}
  {4772--4779} (\bibinfo {year} {2018})}\BibitemShut {NoStop}%
\bibitem [{\citenamefont {Wu}\ \emph {et~al.}(2018)\citenamefont {Wu},
  \citenamefont {Ramsundar}, \citenamefont {Feinberg}, \citenamefont {Gomes},
  \citenamefont {Geniesse}, \citenamefont {Pappu}, \citenamefont {Leswing},\
  and\ \citenamefont {Pande}}]{wu_moleculenet_2018}%
  \BibitemOpen
  \bibfield  {author} {\bibinfo {author} {\bibfnamefont {Z.}~\bibnamefont
  {Wu}}, \bibinfo {author} {\bibfnamefont {B.}~\bibnamefont {Ramsundar}},
  \bibinfo {author} {\bibfnamefont {E.~N.}\ \bibnamefont {Feinberg}}, \bibinfo
  {author} {\bibfnamefont {J.}~\bibnamefont {Gomes}}, \bibinfo {author}
  {\bibfnamefont {C.}~\bibnamefont {Geniesse}}, \bibinfo {author}
  {\bibfnamefont {A.~S.}\ \bibnamefont {Pappu}}, \bibinfo {author}
  {\bibfnamefont {K.}~\bibnamefont {Leswing}}, \ and\ \bibinfo {author}
  {\bibfnamefont {V.}~\bibnamefont {Pande}},\ }\bibfield  {title} {\enquote
  {\bibinfo {title} {{MoleculeNet}: a benchmark for molecular machine
  learning},}\ }\href@noop {} {\bibfield  {journal} {\bibinfo  {journal} {Chem.
  Sci.}\ }\textbf {\bibinfo {volume} {9}},\ \bibinfo {pages} {513--530}
  (\bibinfo {year} {2018})}\BibitemShut {NoStop}%
\bibitem [{\citenamefont {Yao}\ \emph {et~al.}(2018)\citenamefont {Yao},
  \citenamefont {Herr}, \citenamefont {Toth}, \citenamefont {Mckintyre},\ and\
  \citenamefont {Parkhill}}]{yao_tensormol-01_2018}%
  \BibitemOpen
  \bibfield  {author} {\bibinfo {author} {\bibfnamefont {K.}~\bibnamefont
  {Yao}}, \bibinfo {author} {\bibfnamefont {J.~E.}\ \bibnamefont {Herr}},
  \bibinfo {author} {\bibfnamefont {D.~W.}\ \bibnamefont {Toth}}, \bibinfo
  {author} {\bibfnamefont {R.}~\bibnamefont {Mckintyre}}, \ and\ \bibinfo
  {author} {\bibfnamefont {J.}~\bibnamefont {Parkhill}},\ }\bibfield  {title}
  {\enquote {\bibinfo {title} {The {TensorMol}-0.1 model chemistry: a neural
  network augmented with long-range physics},}\ }\href@noop {} {\bibfield
  {journal} {\bibinfo  {journal} {Chem. Sci.}\ }\textbf {\bibinfo {volume}
  {9}},\ \bibinfo {pages} {2261--2269} (\bibinfo {year} {2018})}\BibitemShut
  {NoStop}%
\bibitem [{\citenamefont {Cheng}\ \emph
  {et~al.}(2019{\natexlab{a}})\citenamefont {Cheng}, \citenamefont {Welborn},
  \citenamefont {Christensen},\ and\ \citenamefont
  {Miller~III}}]{cheng_universal_2019}%
  \BibitemOpen
  \bibfield  {author} {\bibinfo {author} {\bibfnamefont {L.}~\bibnamefont
  {Cheng}}, \bibinfo {author} {\bibfnamefont {M.}~\bibnamefont {Welborn}},
  \bibinfo {author} {\bibfnamefont {A.~S.}\ \bibnamefont {Christensen}}, \ and\
  \bibinfo {author} {\bibfnamefont {T.~F.}\ \bibnamefont {Miller~III}},\
  }\bibfield  {title} {\enquote {\bibinfo {title} {A universal density matrix
  functional from molecular orbital-based machine learning: {Transferability}
  across organic molecules},}\ }\href@noop {} {\bibfield  {journal} {\bibinfo
  {journal} {J. Chem. Phys.}\ }\textbf {\bibinfo {volume} {150}},\ \bibinfo
  {pages} {131103} (\bibinfo {year} {2019}{\natexlab{a}})}\BibitemShut
  {NoStop}%
\bibitem [{\citenamefont {Cheng}\ \emph
  {et~al.}(2019{\natexlab{b}})\citenamefont {Cheng}, \citenamefont {Kovachki},
  \citenamefont {Welborn},\ and\ \citenamefont
  {Miller~III}}]{cheng_regression_2019}%
  \BibitemOpen
  \bibfield  {author} {\bibinfo {author} {\bibfnamefont {L.}~\bibnamefont
  {Cheng}}, \bibinfo {author} {\bibfnamefont {N.~B.}\ \bibnamefont {Kovachki}},
  \bibinfo {author} {\bibfnamefont {M.}~\bibnamefont {Welborn}}, \ and\
  \bibinfo {author} {\bibfnamefont {T.~F.}\ \bibnamefont {Miller~III}},\
  }\bibfield  {title} {\enquote {\bibinfo {title} {Regression {Clustering} for
  {Improved} {Accuracy} and {Training} {Costs} with
  {Molecular}-{Orbital}-{Based} {Machine} {Learning}},}\ }\href@noop {}
  {\bibfield  {journal} {\bibinfo  {journal} {J. Chem. Theory Comput.}\
  }\textbf {\bibinfo {volume} {15}},\ \bibinfo {pages} {6668--6677} (\bibinfo
  {year} {2019}{\natexlab{b}})}\BibitemShut {NoStop}%
\bibitem [{\citenamefont {Christensen}, \citenamefont {Faber},\ and\
  \citenamefont {von Lilienfeld}(2019)}]{christensen_operators_2019}%
  \BibitemOpen
  \bibfield  {author} {\bibinfo {author} {\bibfnamefont {A.~S.}\ \bibnamefont
  {Christensen}}, \bibinfo {author} {\bibfnamefont {F.~A.}\ \bibnamefont
  {Faber}}, \ and\ \bibinfo {author} {\bibfnamefont {O.~A.}\ \bibnamefont {von
  Lilienfeld}},\ }\bibfield  {title} {\enquote {\bibinfo {title} {Operators in
  quantum machine learning: {Response} properties in chemical space},}\
  }\href@noop {} {\bibfield  {journal} {\bibinfo  {journal} {J. Chem. Phys.}\
  }\textbf {\bibinfo {volume} {150}},\ \bibinfo {pages} {064105} (\bibinfo
  {year} {2019})}\BibitemShut {NoStop}%
\bibitem [{\citenamefont {Grisafi}\ \emph {et~al.}(2019)\citenamefont
  {Grisafi}, \citenamefont {Fabrizio}, \citenamefont {Meyer}, \citenamefont
  {Wilkins}, \citenamefont {Corminboeuf},\ and\ \citenamefont
  {Ceriotti}}]{grisafi_transferable_2019}%
  \BibitemOpen
  \bibfield  {author} {\bibinfo {author} {\bibfnamefont {A.}~\bibnamefont
  {Grisafi}}, \bibinfo {author} {\bibfnamefont {A.}~\bibnamefont {Fabrizio}},
  \bibinfo {author} {\bibfnamefont {B.}~\bibnamefont {Meyer}}, \bibinfo
  {author} {\bibfnamefont {D.~M.}\ \bibnamefont {Wilkins}}, \bibinfo {author}
  {\bibfnamefont {C.}~\bibnamefont {Corminboeuf}}, \ and\ \bibinfo {author}
  {\bibfnamefont {M.}~\bibnamefont {Ceriotti}},\ }\bibfield  {title} {\enquote
  {\bibinfo {title} {Transferable {Machine}-{Learning} {Model} of the
  {Electron} {Density}},}\ }\href@noop {} {\bibfield  {journal} {\bibinfo
  {journal} {ACS Cent. Sci.}\ }\textbf {\bibinfo {volume} {5}},\ \bibinfo
  {pages} {57--64} (\bibinfo {year} {2019})}\BibitemShut {NoStop}%
\bibitem [{\citenamefont {Smith}\ \emph {et~al.}(2019)\citenamefont {Smith},
  \citenamefont {Nebgen}, \citenamefont {Zubatyuk}, \citenamefont {Lubbers},
  \citenamefont {Devereux}, \citenamefont {Barros}, \citenamefont {Tretiak},
  \citenamefont {Isayev},\ and\ \citenamefont
  {Roitberg}}]{smith_approaching_2019}%
  \BibitemOpen
  \bibfield  {author} {\bibinfo {author} {\bibfnamefont {J.~S.}\ \bibnamefont
  {Smith}}, \bibinfo {author} {\bibfnamefont {B.~T.}\ \bibnamefont {Nebgen}},
  \bibinfo {author} {\bibfnamefont {R.}~\bibnamefont {Zubatyuk}}, \bibinfo
  {author} {\bibfnamefont {N.}~\bibnamefont {Lubbers}}, \bibinfo {author}
  {\bibfnamefont {C.}~\bibnamefont {Devereux}}, \bibinfo {author}
  {\bibfnamefont {K.}~\bibnamefont {Barros}}, \bibinfo {author} {\bibfnamefont
  {S.}~\bibnamefont {Tretiak}}, \bibinfo {author} {\bibfnamefont
  {O.}~\bibnamefont {Isayev}}, \ and\ \bibinfo {author} {\bibfnamefont {A.~E.}\
  \bibnamefont {Roitberg}},\ }\bibfield  {title} {\enquote {\bibinfo {title}
  {Approaching coupled cluster accuracy with a general-purpose neural network
  potential through transfer learning},}\ }\href@noop {} {\bibfield  {journal}
  {\bibinfo  {journal} {Nat. Commun.}\ }\textbf {\bibinfo {volume} {10}},\
  \bibinfo {pages} {2903} (\bibinfo {year} {2019})}\BibitemShut {NoStop}%
\bibitem [{\citenamefont {Unke}\ and\ \citenamefont
  {Meuwly}(2019)}]{unke_physnet_2019}%
  \BibitemOpen
  \bibfield  {author} {\bibinfo {author} {\bibfnamefont {O.~T.}\ \bibnamefont
  {Unke}}\ and\ \bibinfo {author} {\bibfnamefont {M.}~\bibnamefont {Meuwly}},\
  }\bibfield  {title} {\enquote {\bibinfo {title} {{PhysNet}: {A} {Neural}
  {Network} for {Predicting} {Energies}, {Forces}, {Dipole} {Moments} and
  {Partial} {Charges}},}\ }\href@noop {} {\bibfield  {journal} {\bibinfo
  {journal} {J. Chem. Theory Comput.}\ }\textbf {\bibinfo {volume} {15}},\
  \bibinfo {pages} {3678--3693} (\bibinfo {year} {2019})}\BibitemShut {NoStop}%
\bibitem [{\citenamefont {Fabrizio}, \citenamefont {Meyer},\ and\ \citenamefont
  {Corminboeuf}(2020)}]{fabrizio_machine_2020}%
  \BibitemOpen
  \bibfield  {author} {\bibinfo {author} {\bibfnamefont {A.}~\bibnamefont
  {Fabrizio}}, \bibinfo {author} {\bibfnamefont {B.}~\bibnamefont {Meyer}}, \
  and\ \bibinfo {author} {\bibfnamefont {C.}~\bibnamefont {Corminboeuf}},\
  }\bibfield  {title} {\enquote {\bibinfo {title} {Machine learning models of
  the energy curvature vs particle number for optimal tuning of long-range
  corrected functionals},}\ }\href@noop {} {\bibfield  {journal} {\bibinfo
  {journal} {J. Chem. Phys.}\ }\textbf {\bibinfo {volume} {152}},\ \bibinfo
  {pages} {154103} (\bibinfo {year} {2020})}\BibitemShut {NoStop}%
\bibitem [{\citenamefont {Chen}\ \emph {et~al.}(2020)\citenamefont {Chen},
  \citenamefont {Zhang}, \citenamefont {Wang},\ and\ \citenamefont
  {E}}]{chen_ground_2020}%
  \BibitemOpen
  \bibfield  {author} {\bibinfo {author} {\bibfnamefont {Y.}~\bibnamefont
  {Chen}}, \bibinfo {author} {\bibfnamefont {L.}~\bibnamefont {Zhang}},
  \bibinfo {author} {\bibfnamefont {H.}~\bibnamefont {Wang}}, \ and\ \bibinfo
  {author} {\bibfnamefont {W.}~\bibnamefont {E}},\ }\bibfield  {title}
  {\enquote {\bibinfo {title} {Ground state energy functional with
  {Hartree}-{Fock} efficiency and chemical accuracy},}\ }\href@noop {}
  {\bibfield  {journal} {\bibinfo  {journal} {J. Phys. Chem. A}\ }\textbf
  {\bibinfo {volume} {124}},\ \bibinfo {pages} {7155--7165} (\bibinfo {year}
  {2020})}\BibitemShut {NoStop}%
\bibitem [{\citenamefont {Christensen}\ \emph {et~al.}(2020)\citenamefont
  {Christensen}, \citenamefont {Bratholm}, \citenamefont {Faber},\ and\
  \citenamefont {Anatole~von Lilienfeld}}]{christensen_fchl_2020}%
  \BibitemOpen
  \bibfield  {author} {\bibinfo {author} {\bibfnamefont {A.~S.}\ \bibnamefont
  {Christensen}}, \bibinfo {author} {\bibfnamefont {L.~A.}\ \bibnamefont
  {Bratholm}}, \bibinfo {author} {\bibfnamefont {F.~A.}\ \bibnamefont {Faber}},
  \ and\ \bibinfo {author} {\bibfnamefont {O.}~\bibnamefont {Anatole~von
  Lilienfeld}},\ }\bibfield  {title} {\enquote {\bibinfo {title} {{FCHL}
  revisited: {Faster} and more accurate quantum machine learning},}\
  }\href@noop {} {\bibfield  {journal} {\bibinfo  {journal} {J. Chem. Phys.}\
  }\textbf {\bibinfo {volume} {152}},\ \bibinfo {pages} {044107} (\bibinfo
  {year} {2020})}\BibitemShut {NoStop}%
\bibitem [{\citenamefont {Dick}\ and\ \citenamefont
  {Fernandez-Serra}(2020)}]{dick_machine_2020}%
  \BibitemOpen
  \bibfield  {author} {\bibinfo {author} {\bibfnamefont {S.}~\bibnamefont
  {Dick}}\ and\ \bibinfo {author} {\bibfnamefont {M.}~\bibnamefont
  {Fernandez-Serra}},\ }\bibfield  {title} {\enquote {\bibinfo {title} {Machine
  learning accurate exchange and correlation functionals of the electronic
  density},}\ }\href@noop {} {\bibfield  {journal} {\bibinfo  {journal} {Nat.
  Comm.}\ }\textbf {\bibinfo {volume} {11}},\ \bibinfo {pages} {3509} (\bibinfo
  {year} {2020})}\BibitemShut {NoStop}%
\bibitem [{\citenamefont {Liu}\ \emph {et~al.}(2020)\citenamefont {Liu},
  \citenamefont {Lin}, \citenamefont {Jia}, \citenamefont {Cheng},
  \citenamefont {Jiang}, \citenamefont {Guo},\ and\ \citenamefont
  {Ma}}]{liu_transferable_2020}%
  \BibitemOpen
  \bibfield  {author} {\bibinfo {author} {\bibfnamefont {Z.}~\bibnamefont
  {Liu}}, \bibinfo {author} {\bibfnamefont {L.}~\bibnamefont {Lin}}, \bibinfo
  {author} {\bibfnamefont {Q.}~\bibnamefont {Jia}}, \bibinfo {author}
  {\bibfnamefont {Z.}~\bibnamefont {Cheng}}, \bibinfo {author} {\bibfnamefont
  {Y.}~\bibnamefont {Jiang}}, \bibinfo {author} {\bibfnamefont
  {Y.}~\bibnamefont {Guo}}, \ and\ \bibinfo {author} {\bibfnamefont
  {J.}~\bibnamefont {Ma}},\ }\bibfield  {title} {\enquote {\bibinfo {title}
  {Transferable {Multi}-level {Attention} {Neural} {Network} for {Accurate}
  {Prediction} of {Quantum} {Chemistry} {Properties} via {Multi}-task
  {Learning}},}\ }\href {\doibase 10.26434/chemrxiv.12588170.v1} {\  (\bibinfo
  {year} {2020}),\ 10.26434/chemrxiv.12588170.v1}\BibitemShut {NoStop}%
\bibitem [{\citenamefont {Qiao}\ \emph {et~al.}(2020)\citenamefont {Qiao},
  \citenamefont {Welborn}, \citenamefont {Anandkumar}, \citenamefont {Manby},\
  and\ \citenamefont {Miller~III}}]{qiao_orbnet_2020}%
  \BibitemOpen
  \bibfield  {author} {\bibinfo {author} {\bibfnamefont {Z.}~\bibnamefont
  {Qiao}}, \bibinfo {author} {\bibfnamefont {M.}~\bibnamefont {Welborn}},
  \bibinfo {author} {\bibfnamefont {A.}~\bibnamefont {Anandkumar}}, \bibinfo
  {author} {\bibfnamefont {F.~R.}\ \bibnamefont {Manby}}, \ and\ \bibinfo
  {author} {\bibfnamefont {T.~F.}\ \bibnamefont {Miller~III}},\ }\bibfield
  {title} {\enquote {\bibinfo {title} {{OrbNet}: {Deep} {Learning} for
  {Quantum} {Chemistry} {Using} {Symmetry}-{Adapted} {Atomic}-{Orbital}
  {Features}},}\ }\href@noop {} {\bibfield  {journal} {\bibinfo  {journal}
  {arXiv:2007.08026 [physics]}\ } (\bibinfo {year} {2020})}\BibitemShut
  {NoStop}%
\bibitem [{\citenamefont {Manzhos}(2020)}]{manzhos_machine_2020}%
  \BibitemOpen
  \bibfield  {author} {\bibinfo {author} {\bibfnamefont {S.}~\bibnamefont
  {Manzhos}},\ }\bibfield  {title} {\enquote {\bibinfo {title} {Machine
  learning for the solution of the {Schrödinger} equation},}\ }\href@noop {}
  {\bibfield  {journal} {\bibinfo  {journal} {Mach. Learn.: Sci. Technol.}\
  }\textbf {\bibinfo {volume} {1}},\ \bibinfo {pages} {013002} (\bibinfo {year}
  {2020})}\BibitemShut {NoStop}%
\bibitem [{\citenamefont {Nesbet}(1958)}]{nesbet_brueckners_1958}%
  \BibitemOpen
  \bibfield  {author} {\bibinfo {author} {\bibfnamefont {R.~K.}\ \bibnamefont
  {Nesbet}},\ }\bibfield  {title} {\enquote {\bibinfo {title} {Brueckner's
  {Theory} and the {Method} of {Superposition} of {Configurations}},}\
  }\href@noop {} {\bibfield  {journal} {\bibinfo  {journal} {Phys. Rev.}\
  }\textbf {\bibinfo {volume} {109}},\ \bibinfo {pages} {1632--1638} (\bibinfo
  {year} {1958})}\BibitemShut {NoStop}%
\bibitem [{\citenamefont {Møller}\ and\ \citenamefont
  {Plesset}()}]{moller_note_1934}%
  \BibitemOpen
  \bibfield  {author} {\bibinfo {author} {\bibfnamefont {C.}~\bibnamefont
  {Møller}}\ and\ \bibinfo {author} {\bibfnamefont {M.~S.}\ \bibnamefont
  {Plesset}},\ }\bibfield  {title} {\enquote {\bibinfo {title} {Note on an
  {Approximation} {Treatment} for {Many}-{Electron} {Systems}},}\ }\href@noop
  {} {\bibfield  {journal} {\bibinfo  {journal} {Phys. Rev.}\ }\textbf
  {\bibinfo {volume} {46}}}\BibitemShut {NoStop}%
\bibitem [{\citenamefont {Rasmussen}\ and\ \citenamefont
  {Williams}(2006)}]{rasmussen_gaussian_2006}%
  \BibitemOpen
  \bibfield  {author} {\bibinfo {author} {\bibfnamefont {C.~E.}\ \bibnamefont
  {Rasmussen}}\ and\ \bibinfo {author} {\bibfnamefont {C.~K.~I.}\ \bibnamefont
  {Williams}},\ }\href {http://www.gaussianprocess.org/gpml/chapters/RW.pdf}
  {\emph {\bibinfo {title} {Gaussian processes for machine learning}}}\
  (\bibinfo  {publisher} {MIT Press},\ \bibinfo {address} {Cambridge, MA},\
  \bibinfo {year} {2006})\ \bibinfo {note} {publication Title: Gaussian
  processes for machine learning}\BibitemShut {NoStop}%
\bibitem [{\citenamefont {Kapuy}, \citenamefont {Csépes},\ and\ \citenamefont
  {Kozmutza}(1983)}]{kapuy_application_1983}%
  \BibitemOpen
  \bibfield  {author} {\bibinfo {author} {\bibfnamefont {E.}~\bibnamefont
  {Kapuy}}, \bibinfo {author} {\bibfnamefont {Z.}~\bibnamefont {Csépes}}, \
  and\ \bibinfo {author} {\bibfnamefont {C.}~\bibnamefont {Kozmutza}},\
  }\bibfield  {title} {\enquote {\bibinfo {title} {Application of the
  many‐body perturbation theory by using localized orbitals},}\ }\href@noop
  {} {\bibfield  {journal} {\bibinfo  {journal} {Int. J. Quantum Chem.}\
  }\textbf {\bibinfo {volume} {23}},\ \bibinfo {pages} {981} (\bibinfo {year}
  {1983})}\BibitemShut {NoStop}%
\bibitem [{\citenamefont {Bartlett}(1981)}]{bartlett_many-body_1981}%
  \BibitemOpen
  \bibfield  {author} {\bibinfo {author} {\bibfnamefont {R.~J.}\ \bibnamefont
  {Bartlett}},\ }\bibfield  {title} {\enquote {\bibinfo {title} {Many-{Body}
  {Perturbation} {Theory} and {Coupled} {Cluster} {Theory} for {Electron}
  {Correlation} in {Molecules}},}\ }\href@noop {} {\bibfield  {journal}
  {\bibinfo  {journal} {Annu. Rev. Phys. Chem.}\ }\textbf {\bibinfo {volume}
  {32}},\ \bibinfo {pages} {359--401} (\bibinfo {year} {1981})}\BibitemShut
  {NoStop}%
\bibitem [{\citenamefont {Szabo}\ and\ \citenamefont
  {Ostlund}(1996)}]{SzaboSizeConsistent}%
  \BibitemOpen
  \bibfield  {author} {\bibinfo {author} {\bibfnamefont {A.}~\bibnamefont
  {Szabo}}\ and\ \bibinfo {author} {\bibfnamefont {N.~S.}\ \bibnamefont
  {Ostlund}},\ }\href@noop {} {\emph {\bibinfo {title} {{Modern Quantum
  Chemistry}}}}\ (\bibinfo  {publisher} {Dover},\ \bibinfo {address}
  {Mineola},\ \bibinfo {year} {1996})\ pp.\ \bibinfo {pages}
  {261--265}\BibitemShut {NoStop}%
\bibitem [{\citenamefont {Grimme}\ \emph {et~al.}(2016)\citenamefont {Grimme},
  \citenamefont {Hansen}, \citenamefont {Brandenburg},\ and\ \citenamefont
  {Bannwarth}}]{grimme_dispersion_2016}%
  \BibitemOpen
  \bibfield  {author} {\bibinfo {author} {\bibfnamefont {S.}~\bibnamefont
  {Grimme}}, \bibinfo {author} {\bibfnamefont {A.}~\bibnamefont {Hansen}},
  \bibinfo {author} {\bibfnamefont {J.~G.}\ \bibnamefont {Brandenburg}}, \ and\
  \bibinfo {author} {\bibfnamefont {C.}~\bibnamefont {Bannwarth}},\ }\bibfield
  {title} {\enquote {\bibinfo {title} {Dispersion-corrected mean-field
  electronic structure methods},}\ }\href@noop {} {\bibfield  {journal}
  {\bibinfo  {journal} {Chem. Rev.}\ }\textbf {\bibinfo {volume} {116}},\
  \bibinfo {pages} {5105--5154} (\bibinfo {year} {2016})}\BibitemShut {NoStop}%
\bibitem [{\citenamefont {Cheng}\ \emph
  {et~al.}(2019{\natexlab{c}})\citenamefont {Cheng}, \citenamefont {Welborn},
  \citenamefont {Christensen},\ and\ \citenamefont
  {Miller~III}}]{cheng_thermalized_2019}%
  \BibitemOpen
  \bibfield  {author} {\bibinfo {author} {\bibfnamefont {L.}~\bibnamefont
  {Cheng}}, \bibinfo {author} {\bibfnamefont {M.}~\bibnamefont {Welborn}},
  \bibinfo {author} {\bibfnamefont {A.~S.}\ \bibnamefont {Christensen}}, \ and\
  \bibinfo {author} {\bibfnamefont {T.~F.}\ \bibnamefont {Miller~III}},\
  }\bibfield  {title} {\enquote {\bibinfo {title} {Thermalized ({350K}) {QM7b},
  {GDB}-13, water, and short alkane quantum chemistry dataset including
  {MOB}-{ML} features},}\ }\href {https://data.caltech.edu/records/1177} {\
  (\bibinfo {year} {2019}{\natexlab{c}})}\BibitemShut {NoStop}%
\bibitem [{\citenamefont {Burns}\ \emph {et~al.}(2017)\citenamefont {Burns},
  \citenamefont {Faver}, \citenamefont {Zheng}, \citenamefont {Marshall},
  \citenamefont {Smith}, \citenamefont {Vanommeslaeghe}, \citenamefont
  {MacKerell}, \citenamefont {Merz},\ and\ \citenamefont
  {Sherrill}}]{burns_biofragment_2017}%
  \BibitemOpen
  \bibfield  {author} {\bibinfo {author} {\bibfnamefont {L.~A.}\ \bibnamefont
  {Burns}}, \bibinfo {author} {\bibfnamefont {J.~C.}\ \bibnamefont {Faver}},
  \bibinfo {author} {\bibfnamefont {Z.}~\bibnamefont {Zheng}}, \bibinfo
  {author} {\bibfnamefont {M.~S.}\ \bibnamefont {Marshall}}, \bibinfo {author}
  {\bibfnamefont {D.~G.~A.}\ \bibnamefont {Smith}}, \bibinfo {author}
  {\bibfnamefont {K.}~\bibnamefont {Vanommeslaeghe}}, \bibinfo {author}
  {\bibfnamefont {A.~D.}\ \bibnamefont {MacKerell}}, \bibinfo {author}
  {\bibfnamefont {K.~M.}\ \bibnamefont {Merz}}, \ and\ \bibinfo {author}
  {\bibfnamefont {C.~D.}\ \bibnamefont {Sherrill}},\ }\bibfield  {title}
  {\enquote {\bibinfo {title} {The {BioFragment} {Database} ({BFDb}): {An}
  open-data platform for computational chemistry analysis of noncovalent
  interactions},}\ }\href@noop {} {\bibfield  {journal} {\bibinfo  {journal}
  {J. Chem. Phys.}\ }\textbf {\bibinfo {volume} {147}},\ \bibinfo {pages}
  {161727} (\bibinfo {year} {2017})}\BibitemShut {NoStop}%
\bibitem [{\citenamefont {Janet}\ and\ \citenamefont
  {Kulik}(2017)}]{nandy_strategies_2018}%
  \BibitemOpen
  \bibfield  {author} {\bibinfo {author} {\bibfnamefont {J.~P.}\ \bibnamefont
  {Janet}}\ and\ \bibinfo {author} {\bibfnamefont {H.~J.}\ \bibnamefont
  {Kulik}},\ }\bibfield  {title} {\enquote {\bibinfo {title} {Predicting
  electronic structure properties oftransition metal complexes with neural
  networks},}\ }\href@noop {} {\bibfield  {journal} {\bibinfo  {journal} {Chem.
  Sci.}\ }\textbf {\bibinfo {volume} {8}},\ \bibinfo {pages} {5137--5152}
  (\bibinfo {year} {2017})}\BibitemShut {NoStop}%
\bibitem [{cal()}]{caltech_data}%
  \BibitemOpen
  \href@noop {} {}\bibinfo {note} {MOB-ML features, HF energies, pair
  correlation energies, and structures for QM7b-T, GDB-13-T, TM-T, BBI, short
  alkanes, and malonaldehyde will be provided through
  data.caltech.edu.}\BibitemShut {Stop}%
\bibitem [{\citenamefont {Manby}\ \emph {et~al.}(2019)\citenamefont {Manby},
  \citenamefont {Miller~III}, \citenamefont {Bygrave}, \citenamefont {Ding},
  \citenamefont {Dresselhaus}, \citenamefont {Batista-Romero}, \citenamefont
  {Buccheri}, \citenamefont {Bungey}, \citenamefont {Lee}, \citenamefont
  {Meli}, \citenamefont {Miyamoto}, \citenamefont {Steinmann}, \citenamefont
  {Tsuchiya}, \citenamefont {Welborn}, \citenamefont {Wiles},\ and\
  \citenamefont {Williams}}]{manby_entos_2019}%
  \BibitemOpen
  \bibfield  {author} {\bibinfo {author} {\bibfnamefont {F.~R.}\ \bibnamefont
  {Manby}}, \bibinfo {author} {\bibfnamefont {T.~F.}\ \bibnamefont
  {Miller~III}}, \bibinfo {author} {\bibfnamefont {P.}~\bibnamefont {Bygrave}},
  \bibinfo {author} {\bibfnamefont {F.}~\bibnamefont {Ding}}, \bibinfo {author}
  {\bibfnamefont {T.}~\bibnamefont {Dresselhaus}}, \bibinfo {author}
  {\bibfnamefont {F.}~\bibnamefont {Batista-Romero}}, \bibinfo {author}
  {\bibfnamefont {A.}~\bibnamefont {Buccheri}}, \bibinfo {author}
  {\bibfnamefont {C.}~\bibnamefont {Bungey}}, \bibinfo {author} {\bibfnamefont
  {S.~J.~R.}\ \bibnamefont {Lee}}, \bibinfo {author} {\bibfnamefont
  {R.}~\bibnamefont {Meli}}, \bibinfo {author} {\bibfnamefont {K.}~\bibnamefont
  {Miyamoto}}, \bibinfo {author} {\bibfnamefont {C.}~\bibnamefont {Steinmann}},
  \bibinfo {author} {\bibfnamefont {T.}~\bibnamefont {Tsuchiya}}, \bibinfo
  {author} {\bibfnamefont {M.}~\bibnamefont {Welborn}}, \bibinfo {author}
  {\bibfnamefont {T.}~\bibnamefont {Wiles}}, \ and\ \bibinfo {author}
  {\bibfnamefont {Z.}~\bibnamefont {Williams}},\ }\bibfield  {title} {\enquote
  {\bibinfo {title} {entos: {A} {Quantum} {Molecular} {Simulation}
  {Package}},}\ }\href {\doibase 10.26434/chemrxiv.7762646.v2} {\  (\bibinfo
  {year} {2019}),\ 10.26434/chemrxiv.7762646.v2}\BibitemShut {NoStop}%
\bibitem [{\citenamefont {Dunning}(1989)}]{dunning_gaussian_1989}%
  \BibitemOpen
  \bibfield  {author} {\bibinfo {author} {\bibfnamefont {T.~H.}\ \bibnamefont
  {Dunning}},\ }\bibfield  {title} {\enquote {\bibinfo {title} {Gaussian basis
  sets for use in correlated molecular calculations. {I}. {The} atoms boron
  through neon and hydrogen},}\ }\href@noop {} {\bibfield  {journal} {\bibinfo
  {journal} {J. Chem. Phys.}\ }\textbf {\bibinfo {volume} {90}},\ \bibinfo
  {pages} {1007--1023} (\bibinfo {year} {1989})}\BibitemShut {NoStop}%
\bibitem [{\citenamefont {Weigend}\ and\ \citenamefont
  {Ahlrichs}(2005)}]{weigend_balanced_2005}%
  \BibitemOpen
  \bibfield  {author} {\bibinfo {author} {\bibfnamefont {F.}~\bibnamefont
  {Weigend}}\ and\ \bibinfo {author} {\bibfnamefont {R.}~\bibnamefont
  {Ahlrichs}},\ }\bibfield  {title} {\enquote {\bibinfo {title} {Balanced basis
  sets of split valence, triple zeta valence andquadruple zeta valence quality
  for h to rn: Design and assessmentof accuracy},}\ }\href@noop {} {\bibfield
  {journal} {\bibinfo  {journal} {Phys. Chem. Chem. Phys.}\ }\textbf {\bibinfo
  {volume} {7}},\ \bibinfo {pages} {3297--3305} (\bibinfo {year}
  {2005})}\BibitemShut {NoStop}%
\bibitem [{\citenamefont {Weigend}(2002)}]{weigend_fully_2002}%
  \BibitemOpen
  \bibfield  {author} {\bibinfo {author} {\bibfnamefont {F.}~\bibnamefont
  {Weigend}},\ }\bibfield  {title} {\enquote {\bibinfo {title} {A fully direct
  {RI}-{HF} algorithm: {Implementation}, optimised auxiliary basis sets,
  demonstration of accuracy and efficiency},}\ }\href@noop {} {\bibfield
  {journal} {\bibinfo  {journal} {Phys. Chem. Chem. Phys.}\ }\textbf {\bibinfo
  {volume} {4}},\ \bibinfo {pages} {4285--4291} (\bibinfo {year}
  {2002})}\BibitemShut {NoStop}%
\bibitem [{\citenamefont {Weigend}(2006)}]{weigend_accurate_2006}%
  \BibitemOpen
  \bibfield  {author} {\bibinfo {author} {\bibfnamefont {F.}~\bibnamefont
  {Weigend}},\ }\bibfield  {title} {\enquote {\bibinfo {title} {Accurate
  coulomb-fitting basis sets for h to rn},}\ }\href@noop {} {\bibfield
  {journal} {\bibinfo  {journal} {Phys. Chem. Chem. Phys.}\ }\textbf {\bibinfo
  {volume} {8}},\ \bibinfo {pages} {1057--1065} (\bibinfo {year}
  {2006})}\BibitemShut {NoStop}%
\bibitem [{\citenamefont {Boys}(1960)}]{boys_construction_1960}%
  \BibitemOpen
  \bibfield  {author} {\bibinfo {author} {\bibfnamefont {S.~F.}\ \bibnamefont
  {Boys}},\ }\bibfield  {title} {\enquote {\bibinfo {title} {Construction of
  {Some} {Molecular} {Orbitals} to {Be} {Approximately} {Invariant} for
  {Changes} from {One} {Molecule} to {Another}},}\ }\href@noop {} {\bibfield
  {journal} {\bibinfo  {journal} {Rev. Mod. Phys.}\ }\textbf {\bibinfo {volume}
  {32}},\ \bibinfo {pages} {296--299} (\bibinfo {year} {1960})}\BibitemShut
  {NoStop}%
\bibitem [{\citenamefont {Foster}\ and\ \citenamefont
  {Boys}(1960)}]{foster_canonical_1960}%
  \BibitemOpen
  \bibfield  {author} {\bibinfo {author} {\bibfnamefont {J.~M.}\ \bibnamefont
  {Foster}}\ and\ \bibinfo {author} {\bibfnamefont {S.~F.}\ \bibnamefont
  {Boys}},\ }\bibfield  {title} {\enquote {\bibinfo {title} {Canonical
  {Configurational} {Interaction} {Procedure}},}\ }\href@noop {} {\bibfield
  {journal} {\bibinfo  {journal} {Rev. Mod. Phys.}\ }\textbf {\bibinfo {volume}
  {32}},\ \bibinfo {pages} {300--302} (\bibinfo {year} {1960})}\BibitemShut
  {NoStop}%
\bibitem [{\citenamefont {Knizia}(2013)}]{knizia_intrinsic_2013}%
  \BibitemOpen
  \bibfield  {author} {\bibinfo {author} {\bibfnamefont {G.}~\bibnamefont
  {Knizia}},\ }\bibfield  {title} {\enquote {\bibinfo {title} {Intrinsic atomic
  orbitals: an unbiased bridge between quantum theory and chemical concepts},}\
  }\href@noop {} {\bibfield  {journal} {\bibinfo  {journal} {J. Chem. Theory
  Comput.}\ }\textbf {\bibinfo {volume} {9}},\ \bibinfo {pages} {4834}
  (\bibinfo {year} {2013})}\BibitemShut {NoStop}%
\bibitem [{\citenamefont {Werner}\ \emph {et~al.}(2018)\citenamefont {Werner},
  \citenamefont {Knowles}, \citenamefont {Knizia}, \citenamefont {Manby},
  \citenamefont {Schütz},\ and\ \citenamefont
  {{others}}}]{werner_molpro_2018}%
  \BibitemOpen
  \bibfield  {author} {\bibinfo {author} {\bibfnamefont {H.-J.}\ \bibnamefont
  {Werner}}, \bibinfo {author} {\bibfnamefont {P.~J.}\ \bibnamefont {Knowles}},
  \bibinfo {author} {\bibfnamefont {G.}~\bibnamefont {Knizia}}, \bibinfo
  {author} {\bibfnamefont {F.~R.}\ \bibnamefont {Manby}}, \bibinfo {author}
  {\bibfnamefont {M.}~\bibnamefont {Schütz}}, \ and\ \bibinfo {author}
  {\bibnamefont {{others}}},\ }\href@noop {} {\emph {\bibinfo {title}
  {{MOLPRO}, version 2018.3, a package of ab initio programs}}}\ (\bibinfo
  {address} {Cardiff, UK},\ \bibinfo {year} {2018})\BibitemShut {NoStop}%
\bibitem [{\citenamefont {Werner}\ \emph {et~al.}()\citenamefont {Werner},
  \citenamefont {Knowles}, \citenamefont {Manby}, \citenamefont {Black},
  \citenamefont {Doll}, \citenamefont {Hesselmann}, \citenamefont {Kats},
  \citenamefont {Köhn}, \citenamefont {Korona}, \citenamefont {Kreplin},
  \citenamefont {Ma}, \citenamefont {Miller~III}, \citenamefont
  {Mitrushchenkov}, \citenamefont {Peterson}, \citenamefont {Polyak},\ and\
  \citenamefont {Rauhut}}]{werner_molpro_2020}%
  \BibitemOpen
  \bibfield  {author} {\bibinfo {author} {\bibfnamefont {H.-J.}\ \bibnamefont
  {Werner}}, \bibinfo {author} {\bibfnamefont {P.~J.}\ \bibnamefont {Knowles}},
  \bibinfo {author} {\bibfnamefont {F.~B.}\ \bibnamefont {Manby}}, \bibinfo
  {author} {\bibfnamefont {J.~A.}\ \bibnamefont {Black}}, \bibinfo {author}
  {\bibfnamefont {K.}~\bibnamefont {Doll}}, \bibinfo {author} {\bibfnamefont
  {A.}~\bibnamefont {Hesselmann}}, \bibinfo {author} {\bibfnamefont
  {D.}~\bibnamefont {Kats}}, \bibinfo {author} {\bibfnamefont {A.}~\bibnamefont
  {Köhn}}, \bibinfo {author} {\bibfnamefont {T.}~\bibnamefont {Korona}},
  \bibinfo {author} {\bibfnamefont {D.~A.}\ \bibnamefont {Kreplin}}, \bibinfo
  {author} {\bibfnamefont {Q.}~\bibnamefont {Ma}}, \bibinfo {author}
  {\bibfnamefont {T.~F.}\ \bibnamefont {Miller~III}}, \bibinfo {author}
  {\bibfnamefont {A.}~\bibnamefont {Mitrushchenkov}}, \bibinfo {author}
  {\bibnamefont {Peterson}}, \bibinfo {author} {\bibfnamefont {I.}~\bibnamefont
  {Polyak}}, \ and\ \bibinfo {author} {\bibfnamefont {M.}~\bibnamefont
  {Rauhut}, \bibfnamefont {G.~Sibaev}},\ }\bibfield  {title} {\enquote
  {\bibinfo {title} {The molpro quantum chemistry package},}\ }\href@noop {}
  {\bibfield  {journal} {\bibinfo  {journal} {J. Chem. Phys.}\ }\textbf
  {\bibinfo {volume} {152}}}\BibitemShut {NoStop}%
\bibitem [{\citenamefont {Werner}, \citenamefont {Manby},\ and\ \citenamefont
  {Knowles}(2003)}]{werner_fast_2003}%
  \BibitemOpen
  \bibfield  {author} {\bibinfo {author} {\bibfnamefont {H.-J.}\ \bibnamefont
  {Werner}}, \bibinfo {author} {\bibfnamefont {F.~R.}\ \bibnamefont {Manby}}, \
  and\ \bibinfo {author} {\bibfnamefont {P.~J.}\ \bibnamefont {Knowles}},\
  }\bibfield  {title} {\enquote {\bibinfo {title} {Fast linear scaling
  second-order møller-plesset perturbation theory (mp2) using local and
  density fitting approximations},}\ }\href@noop {} {\bibfield  {journal}
  {\bibinfo  {journal} {J. Chem. Phys.}\ }\textbf {\bibinfo {volume} {118}},\
  \bibinfo {pages} {8149} (\bibinfo {year} {2003})}\BibitemShut {NoStop}%
\bibitem [{\citenamefont {Scuseria}\ and\ \citenamefont
  {Lee}(1990)}]{scuseria_comparison_1990}%
  \BibitemOpen
  \bibfield  {author} {\bibinfo {author} {\bibfnamefont {G.~E.}\ \bibnamefont
  {Scuseria}}\ and\ \bibinfo {author} {\bibfnamefont {T.~J.}\ \bibnamefont
  {Lee}},\ }\bibfield  {title} {\enquote {\bibinfo {title} {Comparison of
  coupled‐cluster methods which include the effects of connected triple
  excitations},}\ }\href@noop {} {\bibfield  {journal} {\bibinfo  {journal} {J.
  Chem. Phys.}\ }\textbf {\bibinfo {volume} {93}},\ \bibinfo {pages}
  {5851--5855} (\bibinfo {year} {1990})}\BibitemShut {NoStop}%
\bibitem [{\citenamefont {Hampel}\ and\ \citenamefont
  {Werner}(1996)}]{hampel_local_1996}%
  \BibitemOpen
  \bibfield  {author} {\bibinfo {author} {\bibfnamefont {C.}~\bibnamefont
  {Hampel}}\ and\ \bibinfo {author} {\bibfnamefont {H.~J.}\ \bibnamefont
  {Werner}},\ }\bibfield  {title} {\enquote {\bibinfo {title} {Local treatment
  of electron correlation in coupled cluster theory},}\ }\href@noop {}
  {\bibfield  {journal} {\bibinfo  {journal} {J. Chem. Phys.}\ }\textbf
  {\bibinfo {volume} {104}},\ \bibinfo {pages} {6286--6297} (\bibinfo {year}
  {1996})}\BibitemShut {NoStop}%
\bibitem [{\citenamefont {Schütz}\ and\ \citenamefont
  {Werner}(2000)}]{schutz_local_2000}%
  \BibitemOpen
  \bibfield  {author} {\bibinfo {author} {\bibfnamefont {M.}~\bibnamefont
  {Schütz}}\ and\ \bibinfo {author} {\bibfnamefont {H.-J.}\ \bibnamefont
  {Werner}},\ }\bibfield  {title} {\enquote {\bibinfo {title} {Local
  perturbative triples correction (t) with linear cost scaling},}\ }\href@noop
  {} {\bibfield  {journal} {\bibinfo  {journal} {Chem. Phys. Lett.}\ }\textbf
  {\bibinfo {volume} {218}},\ \bibinfo {pages} {370--378} (\bibinfo {year}
  {2000})}\BibitemShut {NoStop}%
\bibitem [{\citenamefont {Pedregosa}\ \emph {et~al.}(2011)\citenamefont
  {Pedregosa}, \citenamefont {Varoquaux}, \citenamefont {Gramfort},
  \citenamefont {Michel}, \citenamefont {Thirion}, \citenamefont {Grisel},
  \citenamefont {Blondel}, \citenamefont {Prettenhofer}, \citenamefont {Weiss},
  \citenamefont {Dubourg}, \citenamefont {Vanderplas}, \citenamefont {Passos},
  \citenamefont {Cournapeau}, \citenamefont {Brucher}, \citenamefont {Perrot},\
  and\ \citenamefont {Duchesnay}}]{pedregosa_scikit-learn_2011}%
  \BibitemOpen
  \bibfield  {author} {\bibinfo {author} {\bibfnamefont {F.}~\bibnamefont
  {Pedregosa}}, \bibinfo {author} {\bibfnamefont {G.}~\bibnamefont
  {Varoquaux}}, \bibinfo {author} {\bibfnamefont {A.}~\bibnamefont {Gramfort}},
  \bibinfo {author} {\bibfnamefont {V.}~\bibnamefont {Michel}}, \bibinfo
  {author} {\bibfnamefont {B.}~\bibnamefont {Thirion}}, \bibinfo {author}
  {\bibfnamefont {O.}~\bibnamefont {Grisel}}, \bibinfo {author} {\bibfnamefont
  {M.}~\bibnamefont {Blondel}}, \bibinfo {author} {\bibfnamefont
  {P.}~\bibnamefont {Prettenhofer}}, \bibinfo {author} {\bibfnamefont
  {R.}~\bibnamefont {Weiss}}, \bibinfo {author} {\bibfnamefont
  {V.}~\bibnamefont {Dubourg}}, \bibinfo {author} {\bibfnamefont
  {J.}~\bibnamefont {Vanderplas}}, \bibinfo {author} {\bibfnamefont
  {A.}~\bibnamefont {Passos}}, \bibinfo {author} {\bibfnamefont
  {D.}~\bibnamefont {Cournapeau}}, \bibinfo {author} {\bibfnamefont
  {M.}~\bibnamefont {Brucher}}, \bibinfo {author} {\bibfnamefont
  {M.}~\bibnamefont {Perrot}}, \ and\ \bibinfo {author} {\bibfnamefont
  {E.}~\bibnamefont {Duchesnay}},\ }\bibfield  {title} {\enquote {\bibinfo
  {title} {Scikit-learn: machine learning in python (v0.21.2)},}\ }\href@noop
  {} {\bibfield  {journal} {\bibinfo  {journal} {J. Mach. Learn. Res.}\
  }\textbf {\bibinfo {volume} {12}},\ \bibinfo {pages} {2825} (\bibinfo {year}
  {2011})}\BibitemShut {NoStop}%
\bibitem [{\citenamefont {{GPy}}(2012)}]{gpy_gpy_2012}%
  \BibitemOpen
  \bibfield  {author} {\bibinfo {author} {\bibnamefont {{GPy}}},\ }\href
  {http://github.com/SheffieldML/GPy} {\emph {\bibinfo {title} {{GPy}: {A}
  {Gaussian} process framework in python}}}\ (\bibinfo {year}
  {2012})\BibitemShut {NoStop}%
\bibitem [{\citenamefont {Shapeev}\ \emph {et~al.}(2020)\citenamefont
  {Shapeev}, \citenamefont {Gubaev}, \citenamefont {Tsymbalov},\ and\
  \citenamefont {Podryabinkin}}]{shapeev_active_2020}%
  \BibitemOpen
  \bibfield  {author} {\bibinfo {author} {\bibfnamefont {A.}~\bibnamefont
  {Shapeev}}, \bibinfo {author} {\bibfnamefont {K.}~\bibnamefont {Gubaev}},
  \bibinfo {author} {\bibfnamefont {E.}~\bibnamefont {Tsymbalov}}, \ and\
  \bibinfo {author} {\bibfnamefont {E.}~\bibnamefont {Podryabinkin}},\
  }\bibfield  {title} {\enquote {\bibinfo {title} {Active learning and
  uncertainty estimation},}\ }in\ \href@noop {} {\emph {\bibinfo {booktitle}
  {Machine Learning Meets Quantum Physics}}},\ \bibinfo {editor} {edited by\
  \bibinfo {editor} {\bibfnamefont {K.~T.}\ \bibnamefont {Schütt}}, \bibinfo
  {editor} {\bibfnamefont {S.}~\bibnamefont {Chmiela}}, \bibinfo {editor}
  {\bibfnamefont {O.~A.}\ \bibnamefont {von Lilienfeld}}, \bibinfo {editor}
  {\bibfnamefont {A.}~\bibnamefont {Tkatchenko}}, \bibinfo {editor}
  {\bibfnamefont {K.}~\bibnamefont {Tsuda}}, \ and\ \bibinfo {editor}
  {\bibfnamefont {K.-R.}\ \bibnamefont {Müller}}}\ (\bibinfo  {publisher}
  {Springer Nature Switzerland AG.},\ \bibinfo {year} {2020})\ pp.\ \bibinfo
  {pages} {309--329}\BibitemShut {NoStop}%
\bibitem [{\citenamefont {Hirschfeld}\ \emph {et~al.}(2020)\citenamefont
  {Hirschfeld}, \citenamefont {Swanson}, \citenamefont {Yang}, \citenamefont
  {Barzilay},\ and\ \citenamefont {Coley}}]{hirschfeld_uncertainty_2020}%
  \BibitemOpen
  \bibfield  {author} {\bibinfo {author} {\bibfnamefont {L.}~\bibnamefont
  {Hirschfeld}}, \bibinfo {author} {\bibfnamefont {K.}~\bibnamefont {Swanson}},
  \bibinfo {author} {\bibfnamefont {K.}~\bibnamefont {Yang}}, \bibinfo {author}
  {\bibfnamefont {R.}~\bibnamefont {Barzilay}}, \ and\ \bibinfo {author}
  {\bibfnamefont {C.~W.}\ \bibnamefont {Coley}},\ }\bibfield  {title} {\enquote
  {\bibinfo {title} {Uncertainty {Quantification} {Using} {Neural} {Networks}
  for {Molecular} {Property} {Prediction}},}\ }\href@noop {} {\bibfield
  {journal} {\bibinfo  {journal} {J. Chem. Inf. Model}\ }\textbf {\bibinfo
  {volume} {60}},\ \bibinfo {pages} {3770--3780} (\bibinfo {year}
  {2020})}\BibitemShut {NoStop}%
\bibitem [{\citenamefont {Reymond}(2015)}]{reymond_chemical_2015}%
  \BibitemOpen
  \bibfield  {author} {\bibinfo {author} {\bibfnamefont {J.-L.}\ \bibnamefont
  {Reymond}},\ }\bibfield  {title} {\enquote {\bibinfo {title} {The chemical
  space project},}\ }\href@noop {} {\bibfield  {journal} {\bibinfo  {journal}
  {Acc. Chem. Res.}\ }\textbf {\bibinfo {volume} {48}},\ \bibinfo {pages}
  {722--730} (\bibinfo {year} {2015})}\BibitemShut {NoStop}%
\bibitem [{\citenamefont {Blum}\ and\ \citenamefont
  {Reymond}(2009)}]{blum_970_2009}%
  \BibitemOpen
  \bibfield  {author} {\bibinfo {author} {\bibfnamefont {L.~C.}\ \bibnamefont
  {Blum}}\ and\ \bibinfo {author} {\bibfnamefont {J.-L.}\ \bibnamefont
  {Reymond}},\ }\bibfield  {title} {\enquote {\bibinfo {title} {970 {Million}
  {Druglike} {Small} {Molecules} for {Virtual} {Screening} in the {Chemical}
  {Universe} {Database} {GDB}-13},}\ }\href@noop {} {\bibfield  {journal}
  {\bibinfo  {journal} {J. Am. Chem. Soc.}\ }\textbf {\bibinfo {volume}
  {131}},\ \bibinfo {pages} {8732} (\bibinfo {year} {2009})}\BibitemShut
  {NoStop}%
\bibitem [{\citenamefont {Weymuth}\ \emph {et~al.}(2014)\citenamefont
  {Weymuth}, \citenamefont {Couzijn}, \citenamefont {Chen},\ and\ \citenamefont
  {Reiher}}]{weymuth_new_2014}%
  \BibitemOpen
  \bibfield  {author} {\bibinfo {author} {\bibfnamefont {T.}~\bibnamefont
  {Weymuth}}, \bibinfo {author} {\bibfnamefont {E.~P.~A.}\ \bibnamefont
  {Couzijn}}, \bibinfo {author} {\bibfnamefont {P.}~\bibnamefont {Chen}}, \
  and\ \bibinfo {author} {\bibfnamefont {M.}~\bibnamefont {Reiher}},\
  }\bibfield  {title} {\enquote {\bibinfo {title} {New {Benchmark} {Set} of
  {Transition}-{Metal} {Coordination} {Reactions} for the {Assessment} of
  {Density} {Functionals}},}\ }\href@noop {} {\bibfield  {journal} {\bibinfo
  {journal} {J. Chem. Theory ut.}\ }\textbf {\bibinfo {volume} {10}},\ \bibinfo
  {pages} {3092--3103} (\bibinfo {year} {2014})}\BibitemShut {NoStop}%
\bibitem [{\citenamefont {Husch}, \citenamefont {Freitag},\ and\ \citenamefont
  {Reiher}(2018)}]{husch_calculation_2018}%
  \BibitemOpen
  \bibfield  {author} {\bibinfo {author} {\bibfnamefont {T.}~\bibnamefont
  {Husch}}, \bibinfo {author} {\bibfnamefont {L.}~\bibnamefont {Freitag}}, \
  and\ \bibinfo {author} {\bibfnamefont {M.}~\bibnamefont {Reiher}},\
  }\bibfield  {title} {\enquote {\bibinfo {title} {Calculation of {Ligand}
  {Dissociation} {Energies} in {Large} {Transition}-{Metal} {Complexes}},}\
  }\href@noop {} {\bibfield  {journal} {\bibinfo  {journal} {J. Chem. Theory
  Comput.}\ }\textbf {\bibinfo {volume} {14}},\ \bibinfo {pages} {2456--2468}
  (\bibinfo {year} {2018})}\BibitemShut {NoStop}%
\bibitem [{\citenamefont {Bartlett}\ and\ \citenamefont
  {Ranasinghe}(2017)}]{bartlett_power_2017}%
  \BibitemOpen
  \bibfield  {author} {\bibinfo {author} {\bibfnamefont {R.~J.}\ \bibnamefont
  {Bartlett}}\ and\ \bibinfo {author} {\bibfnamefont {D.~S.}\ \bibnamefont
  {Ranasinghe}},\ }\bibfield  {title} {\enquote {\bibinfo {title} {The power of
  exact conditions in electronic structure theory},}\ }\href@noop {} {\bibfield
   {journal} {\bibinfo  {journal} {Chem. Phys. Lett.}\ }\textbf {\bibinfo
  {volume} {669}},\ \bibinfo {pages} {54--70} (\bibinfo {year}
  {2017})}\BibitemShut {NoStop}%
\end{thebibliography}%
\end{document}